\documentclass[%
 reprint,
 amsmath,amssymb,
 aps,
]{revtex4-2}

\usepackage{graphicx}
\usepackage{dcolumn}
\usepackage{bm}
\usepackage{graphicx}	
\usepackage{amsmath}	
\usepackage{amssymb}	
\usepackage{algpseudocode}
\usepackage{float}
\usepackage[normalem]{ulem}
\usepackage{color}

\usepackage{hyperref}

\definecolor{MyBlue}{rgb}{0.15,0.15,0.70}
\hypersetup{
colorlinks=true,
citecolor=MyBlue,
linkcolor=MyBlue,
urlcolor=MyBlue
}

\begin{document}

\preprint{APS/123-QED}

\title{Seeking New Physics in Cosmology with Bayesian Neural Networks:\\ Dark Energy and Modified Gravity}

\author{M. Mancarella, J. Kennedy, B. Bose, L. Lombriser}
\affiliation{%
 D\'epartement de Physique Th\'eorique, Universit\'e de Gen\`eve, 24~quai Ernest Ansermet, 1211~Gen\`eve~4, Switzerland
}%

\date{\today}

\begin{abstract}
We study the potential of Bayesian Neural Networks (BNNs) to detect new physics in the dark matter power spectrum, concentrating here on evolving dark energy and modifications to General Relativity. 
After introducing a new technique to quantify classification uncertainty in BNNs, we train two BNNs on mock matter power spectra produced using the publicly available code $\tt{ReACT}$ in the $k$-range $\left(0.01 - 2.5\right)  \, h \mathrm{Mpc}^{-1} $ and redshift bins $\left(0.1,0.478,0.783,1.5\right)$ with Euclid-like noise. 
The first network classifies spectra into five labels including $\Lambda$CDM, $f(R)$, $w$CDM, Dvali-Gabadaze-Porrati (DGP) gravity and a ``random'' class whereas the second is trained to distinguish $\Lambda$CDM from non-$\Lambda$CDM.
Both networks achieve a comparable training, validation and test accuracy of $\sim 95\%$.
Each network is also capable of correctly classifying spectra with deviations from $\Lambda$CDM that were not included in the training set, demonstrated with spectra generated using the growth-index $\gamma$.  
To obtain an indication of the BNNs classification capability, we compute the smallest deviation from $\Lambda$CDM such that the noise-averaged non-$\Lambda$CDM classification probability is at least $95\%$ according to our estimated error quantification, finding these bounds to be $f_{R0} \lesssim 10^{-7}$, $\Omega_{rc} \lesssim 10^{-2} $, $-1.05 \lesssim w_0 \lesssim 0.95 $, $-0.2 \lesssim w_a \lesssim 0.2 $, $0.52 \lesssim \gamma \lesssim 0.59 $. 
The bounds on $f(R)$ can be improved by training a specialist network to distinguish solely between $\Lambda$CDM and $f(R)$ power spectra which can detect a non-zero $f_{R0}$ at $\mathcal{O}\left(10^{-8}\right)$.
We expect that further developments, such as the inclusion of smaller length scales or additional extensions to $\Lambda$CDM, will only improve the potential of BNNs to indicate the presence of new physics in cosmological datasets, regardless of the underlying theory.  

\end{abstract}

\maketitle




\section{Introduction}

The scientific method is based upon the meticulous comparison of theoretical hypotheses with observations. 
A hypothesis can be promoted to a foundational theory once it has rigorously 
satisfied a multitude of observational tests. 
Such is the case with the concordance cosmological model $\Lambda$CDM, named after the two dominant components that contribute to the current energy density of the Universe: the cosmological constant ($\Lambda$) and cold dark matter (CDM). 
Despite their dominant contribution to the stress-energy density of the Universe, the fundamental nature of both dark matter and dark energy remains a mystery. 
Determining the physical nature of these two components is a central challenge in modern physics. 
Combined with the task of furthering our understanding of dark energy and dark matter is the requirement to test Einstein's theory of General Relativity (GR) in the hitherto un-explored cosmological regime.
Potential modifications to gravitational dynamics at cosmological length scales may also play a part in providing an explanation for dark energy and dark matter.     
A principal pursuit of contemporary cosmology is therefore to stringently compare both GR and $\Lambda$CDM against a considerable collection of alternative models \cite{Copeland:2006wr, Clifton:2011jh, Joyce:2014kja, Bull:2015stt, Koyama:2015vza, Joyce:2016vqv, Ishak:2018his}. 
The Large-Scale Structure (LSS) of the Universe provides an ideal testing ground for competing hypotheses. 
Assuming that CDM can be treated as a perfect fluid, it undergoes gravitational collapse into localised over-densities, generating gravitational wells which the galaxy distribution subsequently traces. 
By correlating galaxy positions over a large volume, a statistical description of how the underlying dark matter clusters can be obtained.
This is largely characterised by the two-point correlation function or the power spectrum in Fourier space. 
Although the distribution of dark matter is not directly observable, modern cosmological surveys use observables such as galaxy clustering \cite{Alam:2016hwk} or weak lensing \cite{Asgari:2020wuj} to probe the underlying dark matter distribution.
The next generation of galaxy surveys such as Euclid \cite{Laureijs:2011gra} and LSST \cite{Ivezic:2008fe} have the capability to measure the cosmological galaxy distribution with extremely high precision, especially at length scales where the cosmological background becomes subdominant to baryonic and nonlinear gravitational physics. 
Analytic methods are impractical in this regime as the evolution equations do not possess closed-form solutions.   
Cosmological N-body simulations can provide highly accurate numerical predictions, yet their computational cost renders them unsuitable for constraining model parameters in Markov Chain Monte Carlo (MCMC) analyses. 
Motivated by this issue, Refs.~\cite{Knabenhans:2018cng, Takahashi:2012em, Mead:2015yca} constructed emulators and nonlinear models for the matter power spectrum, with extensions to include deviations from $\Lambda$CDM developed in Refs.~\cite{Winther:2019mus, Mead:2016zqy, Zhao:2013dza}, ~\cite{Mead:2016zqy}. 
These are fast and accurate methods which compute predictions for the shape of the matter power spectrum but are limited to an underlying hypothesis. 
Recently, Ref.~\cite{Cataneo:2018cic} provided a method to predict the shape of the nonlinear matter power spectrum for a wide range of models which was subsequently implemented into a code called ${\tt ReACT}$ in Ref.~\cite{Bose:2020wch}. 
Using this framework it is possible to generate a large dataset of mock matter power spectra for a broader class of extensions to $\Lambda$CDM with varying values of the model parameters.   

Such tools enable one to extract information from a large range of length scales, substantially improving the constraining power.
MCMC analyses are frequently employed to determine whether physics beyond $\Lambda$CDM is present in cosmological data.
To consistently constrain beyond $\Lambda$CDM physics in such analyses, one must choose a finite set of parameters quantifying the new physics.
It turns out that the number of parameters needed to do this including nonlinear scales while remaining agnostic to the underlying fundamental physics is immense (see Ref.~\cite{Kennedy:2019nie} for example), making such an analysis currently unfeasible.
Consequently, current analyses either restrict themselves to the linear regime of structure formation or perform a model-by-model analysis.
It is worth noting that even if computational expense was not an issue, the simple inclusion of such a large additional parameter space would strongly penalise the extended-modelling on the basis of the Bayesian evidence.
It is therefore of interest to examine alternative approaches which do not rely on picking an effective set of parameters and are less computationally expensive.

With the ability to produce a large dataset consisting of power spectra for a variety of models, it is natural to consider the capability of Deep Neural Networks (DNNs) to classify power spectra according to their underlying cosmological model.
However, the prediction given by a trained DNN can be subject to several sources of uncertainty. 
Adding a slight perturbation to the input, passing the input to a network with a different architecture or training on a separate subset of the training set could all drastically alter the result \cite{2013arXiv1312.6199S, 2014arXiv1412.1897N, gal2016dropout}.
Taking these issues into account is therefore crucial to obtaining statistically robust predictions with neural networks.    
Quantifying the potential variability of the prediction, and in turn the confidence, is extremely difficult with DNNs. %

Bayesian Neural Networks (BNNs) try to model the uncertainty by replacing each weight in the network by a distribution initialised to a prior \cite{MacKay, Neal, Blundell2015,  gal2016dropout, Charnock2020, jospin2020handson}.
Rather than obtaining the same point-like output for an example with every pass through the network, the BNN's prediction varies as each pass draws a different sample from the weight distribution.
By repeatedly passing an example to the BNN, one obtains a distribution of predictions conditioned on the training data, the network architecture, the noise in the data along with other potentially unknown sources of uncertainty.
A quantitative estimate of the classification uncertainty can be obtained in minutes once the BNN has been trained.
As an additional advantage, BNNs naturally provide a regularisation procedure for preventing overfitting \cite{blundell2015weight}.
BNNs have recently been applied in many fields such as gravitational waves \cite{Shen:2019vep, Lin:2020aps, Killestein:2021oqb}, the Cosmic Microwave Background \cite{2020PhRvD.102j3509H, hortua2020accelerating}, autonomous driving \cite{2019arXiv190909884M}, cellular image classification \cite{Deodato824862} and the detection and classification of supernovae \cite{Ramanah:2021bpb, 2020MNRAS.491.4277M}.

In this paper we explore the potential of BNNs to classify non-$\Lambda$CDM models from the matter power spectrum. 
In particular, BNNs can be trained on as many deviations from $\Lambda$CDM as can be implemented in numerical codes such as {\tt ReACT}. 
Even if none of these theories turn out to be the correct model they are all representative of possible sources of new physics. 
We will therefore investigate whether BNNs can identify general deviations from concordance cosmology based on the observational features of known models.

The goal is, at the very least, to develop a promising tool to inform standard and more rigorous MCMC analyses by providing a refinement of the theoretical parameter space that needs to be explored. On the other hand, the possibility of constructing a well-defined probability distribution which accounts for all sources of uncertainty in the prediction from a BNN is an open research question~\cite{Mackay1995,  MacKay1998, Graves2011, lakshminarayanan2017simple, depeweg2018decomposition, Ritter2018, shridhar2019uncertainty, wu2019quantifying, ovadia2019trust, hobbhahn2020fast}. Should this become possible, this method could be promoted to a statistical tool competitive to MCMC. 
Regardless of this possibility, DNNs can simply be used as a tool to compress the information from the power spectrum in a small set of numbers, that can be in turn combined with other ML-based methods in rigorous statistical frameworks (such as Approximate Bayesian Computation, ABC \cite{fan2018abc}, or Likelihood-Free Inference \cite{Alsing:2019xrx}, LFI) to perform model selection in a fully ML-based fashion. 
In any case, it it worthwhile to assess their potential, and this work is a first step in this direction.
\begin{figure}
\centering
\includegraphics[width=.49\textwidth]{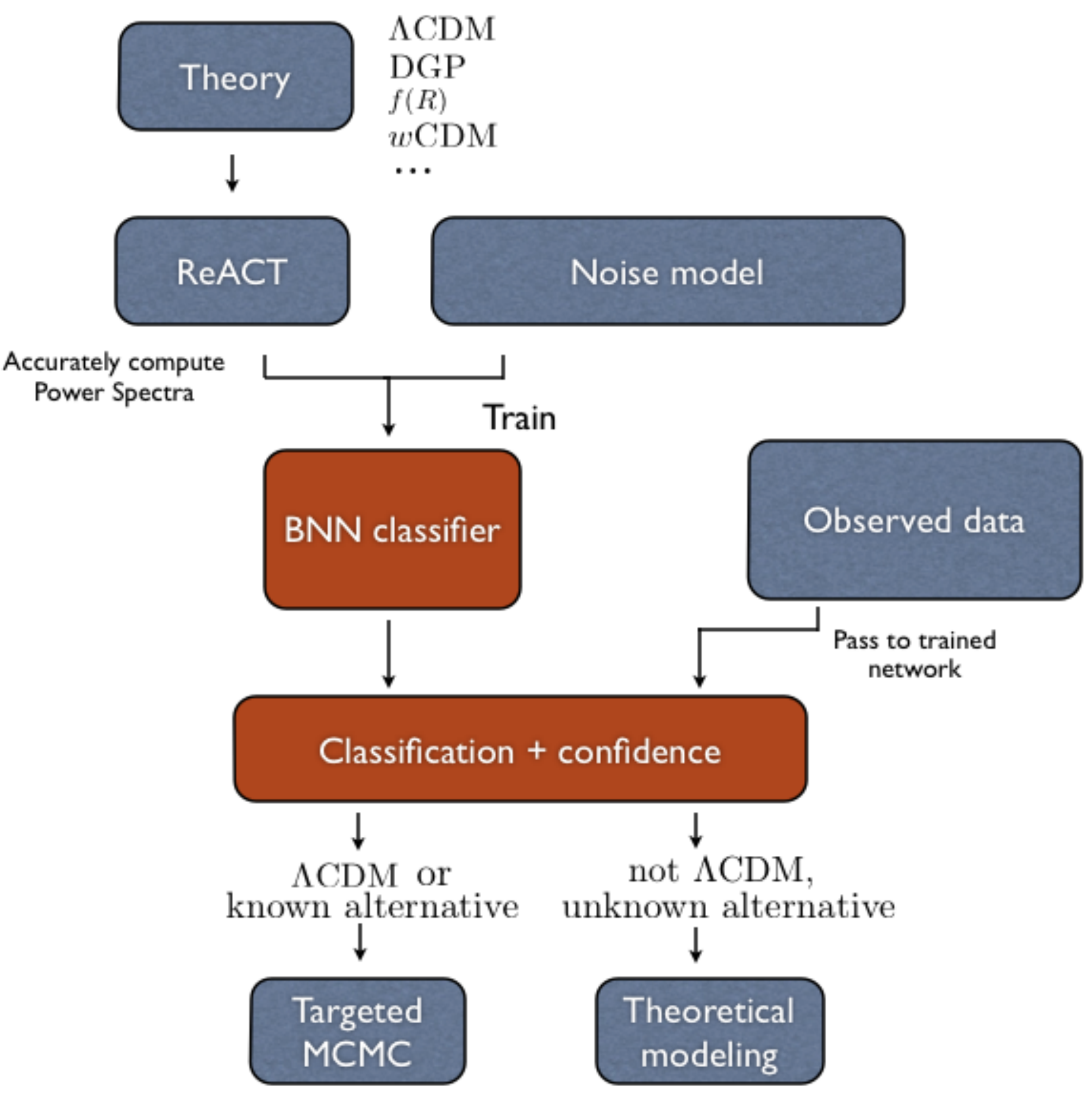}
\caption{ Representation of the workflow presented in the paper to study the presence of deviations from $\Lambda$CDM in the matter power spectrum. The key elements of the method, namely the use of a Bayesian Neural Network and a novel way to quantify the confidence, are colored in red. 
}
\label{fig:workflow}
\end{figure}

In Fig.~\ref{fig:workflow} we show a schematic representation of the method.
Using ${\tt ReACT}$ to generate a training set of thousands of example matter power spectra for both $\Lambda$CDM and selected extensions, 
we train two BNNs to classify the spectra according to the underlying model.
A five-label BNN is trained to classify an example spectrum as either $\Lambda$CDM or one of four chosen extensions while a two-label network is trained simply to classify between $\Lambda$CDM and non-$\Lambda$CDM.
Following the introduction of a novel method to construct a well-defined probability distribution from the output of a BNN in order to take into account the effect of the uncertainty in the final classification (thus preventing the network from being overconfident), we evaluate the performance of each network on the training, validation and test sets. 

In addition, we determine the minimal deviation in the model parameters for each chosen non-$\Lambda$CDM model such that the five-label BNN classifies them as non-$\Lambda$CDM with some specified probability before passing the same spectra through the two-label BNN to compare their performance.    
After studying how effective each BNN is at recognising spectra which do not belong to any class in the training set, we compare their predicted classification probabilities averaged over noise realisations for different values of the model parameters. 
Finally, we examine the potential benefits of training specialist networks on selected subsets of the original classes in the training set. 
While we only consider three well studied dark energy and modified gravity models in this work, this method can be extended to general extensions to $\Lambda$CDM such as massive neutrinos and Horndeski scalar-tensor theory as long as rapid and accurate predictions for the shape of the matter power spectrum can be computed. The effect of neutrino masses and baryonic feedback can now be taken into account and will be the subject of future work, given that it has been recently implemented in {\tt ReACT} \cite{Bose:2021mkz}.

This paper is laid out as follows. 
Sec.~\ref{sec:BNNs} presents a concise theoretical background to both DNNs and BNNs.  
Sec.~\ref{sec:training} then describes the generation and preparation of the training, validation and test data before they can be passed to the BNN, followed by a discussion of the BNN's architecture.
Sec.~\ref{sec:results} discusses the overall performance of each BNN. 
We determine the values of the model parameters in each non-$\Lambda$CDM model such that the five and two-label BNN's are confident that a spectrum deviates from $\Lambda$CDM.
After analysing how sensitive each classification was to the noise in the power spectrum we also discuss the notion of specialist networks.
Sec.~\ref{sec:outlook} lays out potential avenues that should be explored in future studies before we conclude in Sec.~\ref{sec:conclusions}.


\section{Bayesian Neural Networks}\label{sec:BNNs}

Neural networks are becoming ever more widely employed in physics. The interested reader can find a review of the core concepts surrounding the use of neural networks in supervised classification problems in App.~\ref{app:NNintro}, to which we refer for the basic concepts used in the rest of the paper. For a thorough treatment see Refs.~\cite{Goodfellow-et-al-2016,Mehta:2018dln,Carleo:2019ptp}. In this section we discuss the aspects of BNNs relevant to this work, in particular the quantification of the classification uncertainty.

\subsection{Uncertainty in BNN classifiers} 
\label{sec:BNN_uncertainty}

Two principal sources of uncertainty can be identified in the prediction given by a trained network for a new example, namely \emph{aleatoric} and \emph{epistemic} uncertainty \cite{KIUREGHIAN2009105}.
The former encompasses any intrinsic, non-reducible, uncertainty due to the stochasticity of the observations, while the latter describes uncertainty in the model.
High epistemic uncertainty quantifies limitations in modelling accuracy and could be reduced by, for example, choosing a more appropriate model architecture, adjusting the hyperparameters, or training with more data. 

Differently from traditional NNs, BNNs can give an estimate of both sources of uncertainty. Technical details, including a discussion on the difference with traditional NNs, are given in App.~\ref{app:BNNtrain}.  The key concept is the replacement of the networks' weights $w$ with distributions, so that an approximate posterior distribution of the weights, $q_{\theta}(w)$, can be learned by variational inference, instead of learning a single value for each weight as in traditional NNs.
This means that after training the learned distribution can be used to obtain predictions marginalised over the weights, rather than point-like prediction. This in turn allows to take into account potential variability in the network's output and the relative uncertainty as we will now show.

In the case of a classification problem with $N$ classes, the final layer of the network outputs an $N$-dimensional vector with components that sum to one and can therefore be interpreted as probabilities. We denote these components by $p(y_{i}^{\star}=1 | X^{\star},  w, \mathcal{D})$ ($i\in \{1, ..., N\}$) for a new example with features $X^{\star}$ and one-hot encoded label $y^{\star}$, for a given realisation of the weights $w$ and conditioned on the training data $\mathcal{D}$. %
Marginalisation over the weights can be obtained via Monte Carlo sampling from $q_{\theta}(w)$, giving for each component of the one-hot encoded label vector
\begin{align}\label{probabilityBayes}
    &\mu_{i} \equiv p(y_{i}^{\star}=1 | X^{\star}, \mathcal{D}) \approx  \frac{1}{N_{S}} \sum_{\alpha=1}^{N_{S}}  \, p_{\alpha} , \\ \nonumber
    & p_{\alpha} \equiv p(y_{i}^{\star}=1 | X^{\star},  w_\alpha, \mathcal{D})\, , \quad w_{\alpha} \sim q_{\theta}(w | \mathcal{D}) \, ,
\end{align}
where $N_{S}$ is the number of samples and throughout this paper we use Greek indices to denote MC samples and Latin indices to denote vector components.
Eq.~\ref{probabilityBayes} is the Monte Carlo approximation of the exact expression in Eq.~\ref{probabilityBayesExact}.
A prediction for the label for a new example with features $X^{\star}$ is obtained by assigning the label to the maximum output probability
$y_{\text{pred}}^{\star} = {\arg\max}_i \, \mu_{i}$ $(i=1...N)$, if this exceeds a chosen threshold probability $p_{th}$.

Defining $\mu$ to be the vector with components $\mu_i$ from Eq.~\eqref{probabilityBayes}, the full covariance of the classification is given by \cite{KWON2020106816}
\begin{align}\label{fullCov}
    \Sigma_{q_{\theta}} &=
    {\mathbb{E}}_{q_{\theta}} \Big[ Cov_{ p(y_{i}^{\star}=1 | X^{\star}, w, \mathcal{D})}(y^{\star}) \Big] \nonumber \\ & +Cov_{q_{\theta}}\Big[ {\mathbb{E}}_{ p(y_{i}^{\star}=1 | X^{\star}, w, \mathcal{D})}(y^{\star}) \Big] \, \nonumber \\
    &=    \frac{1}{N_{S}}\sum_{\alpha=1}^{N_{S}} \, \left( \text{diag}(p_\alpha) -   p_\alpha^{\otimes 2}  \right) + \frac{1}{N_{S}}\sum_{\alpha=1}^{N_{S}} (p_\alpha-\mu)^{\otimes 2} \, \nonumber   \\  
    & =  \text{diag} ( {\mu} )- { {\mu} }^{\otimes 2}\, ,
\end{align}
where the first line follows from the definition of the covariance
and the second from the use of Eq.~\eqref{probabilityBayes} with the following property of a multinomial distribution (which is used as the likelihood of the optimisation problem as customary in classification tasks, see App.~\ref{app:BNNtrain}): ${\mathbb{E}}_{ p(y_{i}^{\star}=1 | X^{\star}, w, \mathcal{D})}(y^{\star}) = p(y_{i}^{\star}=1 | X^{\star}, w, \mathcal{D})$. 
This shows that the covariance is simply the standard multinomial covariance over the distribution of MC averages $\mu$. 
The second term in the sum is the standard mean-squared error coming from the fact that the weights have a distribution $q_{\theta}(w | \mathcal{D})$, hence it corresponds to the epistemic uncertainty. The first term encodes the contribution to the variance marginalising over $q_{\theta}(w | \mathcal{D})$, and as such it describes the aleatoric uncertainty.
In order not to yield over-confident estimates of whether a given power spectrum is classified as $\Lambda$CDM or not it is important to accommodate both sources of uncertainty into the analysis. 
When training on data coming from real-world observations, one has no means to reduce the aleatoric uncertainty (this is why this is sometimes referred to as ``uncertainty in the data''). In this paper we train a network on simulated noisy data, as described in \ref{sec:datagen}. In principle, the knowledge of the model from which the noise is drawn could be incorporated in the loss. Here we rather make the choice of treating noise as an effective aleatoric uncertainty, and including its effect in the classification uncertainty. Of course, a dependence on the noise model will be inherited during training. We note however that any data analysis tool relies on a model of the noise.

In order to compute the uncertainty, it must be kept in mind that despite $\mu_{i}$ being a probability by construction, it still does not represent an inferred ``true probability'' for the resultant classification as occurs in a likelihood or MCMC analysis.
The quantity $\mu_{i}$ should rather be interpreted as a parameter in itself used to classify a given spectrum if the magnitude exceeds the chosen threshold probability $p_{th}$.
Constructing a confidence in the classification at test time requires a joint distribution on $\mu_{i}$ to compute the sub-volume where $\mu_{i}>p_{th}$. 
We shall detail in the following subsection how we utilise the uncertainty in Eq.~\eqref{fullCov} to estimate the confidence in a particular classification. 
We stress here that while it is tempting to view BNNs as being able to provide a clear and statistically rigorous definition of probability in the classification, we should keep in mind that the model of the error is still subject to approximations, such as the variational approach described in this section and the choice of the parametric distribution $q_{\theta}(w)$.
For these reasons, it is also important to point out that the use of the definition ``Bayesian'' Neural Networks in the formulation used in this work comes from the Machine Learning literature, and the estimated classification probabilities should not be confused with the result of a truly Bayesian model selection as resulting, for example, from the computation of Bayesian evidences with a nested sampling algorithm.
Rather, at the current state of the art BNNs should be viewed as tools that at least enable one to introduce a model of the uncertainty, preventing overly optimistic interpretations of the results as well as providing an effective regularisation procedure.

\subsection{Quantifying the classification confidence} 
\label{sec:construction_conf}

Currently there is no well-established method of quantifying the confidence in a prediction from a BNN. 
In general, obtaining a classification confidence requires the definition of a probability distribution over the softmax output of the network. 
One possibility is the Dirichlet distribution which is both analytic and possesses a natural interpretation as a distribution over probabilities being defined on the $N$-simplex. 
Possible approaches include mapping the variance of the pre-softmax network output to Dirchlet parameters \cite{hobbhahn2020fast}, directly including a Dirichlet distribution in the loss function definition \cite{wu2019quantifying}, or training ``prior networks'' that directly output the parameters of the Dirichlet distribution \cite{malinin2018predictive}. 
Another approach is to empirically define a ``confidence score'' using $\mu$ and the covariance in Eq.~\eqref{fullCov} \cite{Deodato824862,Lin:2020aps}.

In this work we introduce a novel approach which also directly utilises the covariance in Eq.~\eqref{fullCov}. 
We consider a random variable $x \in \mathbb{R}^N$ distributed as a multivariate Gaussian truncated to lie between 0 and 1 with mean $\mu$ and covariance $\Sigma_{q_{\theta}}$ and compute the volume where $x_i>p_{\text{th}}$ $\forall i=1...N$ to obtain the confidence. 
In practice, the definition of such a distribution is complicated by the fact that the components $x_i$ are not independent as both they and the means $\mu_i$ must sum to one. 
This interdependency of the components $x_{i}$ implies one cannot define a multivariate Gaussian directly with the covariance Eq.~\eqref{fullCov}. 
The full derivation of the resulting probability distribution we denote as $\mathcal{F}(x; \mu, \Sigma_{q_{\theta}})$ is outlined in App.~\ref{app:prob} with the final result being
\begin{align}\label{probDist}
\mathcal{F}(x; \mu, \Sigma_{q_{\theta}}) & = \,  \delta\Big( 1-\sum_{j=1}^N x_j \Big)  \times \sqrt{N} \\ \nonumber
& \times \prod_{i=1}^{N-1} {\mathcal{\tilde{N}}} \Big( \left[{B}^{-1} (x-\mu)\right]_i; 0, \left[{B}^{-1} \Sigma_{q_{\theta}}{B}\right]_{ii} \Big) \, ,
\end{align}
where $B$ is the matrix which diagonalises $ \Sigma_{q_{\theta}}$ and $\mathcal{\tilde{N}}$ denotes a Gaussian truncated between 0 and 1. 
The Dirac delta-function enforces the constraint that the components must sum to one with the remaining terms being the product of $N-1$ one-dimensional Gaussians each with a variance given by the non-null eigenvalues of $\Sigma_{q_{\theta}}$.
By using the threshold probability $p_{th}$ and marginalising over the remaining labels, the probability an example is assigned the label $I$ can then be defined as 
\begin{equation}\label{pGauss}
    P_I \equiv \int_{p_{\text{th}}}^1 dx_I\, \int_0^1 dx_1...\hat{dx_I}...dx_N\, \mathcal{F}(x; \mu, \Sigma_{q_{\theta}}) \, ,
\end{equation}
where $\hat{dx_I}$ denotes that the integration on the $I$-th variable is omitted. 
In practice, to compute the integrals in Eq.~\eqref{pGauss} we sample Eq.~\eqref{probDist} as outlined in App.~\ref{app:prob} and determine the fraction of samples which satisfy $x_I>p_{\text{th}}$. 
If no components of a sample exceed $p_{th}$ then it is not assigned a label and the total fraction of such samples gives the probability the example is un-classifiable.

The probability $P_{I}$ encodes an estimate of the uncertainty in the classification and can be used to construct a first approximation of the confidence in the following manner. 
Denoting $P_{\text{gauss}}(n\sigma)$ to be the usual volume of a Gaussian distribution in the interval centered on a mean value with width $n\times \sigma$ we define there to be a $n\sigma$ detection of a deviation from $\Lambda$CDM if $P_{\Lambda CDM}=1-P_{\text{gauss}}(n\sigma)$.
For example, a $2\sigma$ detection corresponds to $P_{\Lambda CDM}=1-P_{\text{gauss}}(2\sigma)=1-0.9545=0.0455$. 
Moreover, if an example is classified with the label $I$ at less than $1\sigma$ confidence such that $P_I<0.68$, we shall not consider this a detection even if $\mu_I>p_{\text{th}}$. 
Note that $1-P_{\Lambda CDM}$ represents the probability of an example not being $\Lambda$CDM, including the probability of the example being un-classifiable.
It therefore does not strictly represent the probability of a non-$\Lambda$CDM detection but also includes the probability that the BNN isn't able to determine which class from the training set the example belongs to.


 %
 \begin{table*}
\centering
\begin{tabular}{| c | c | c | c | c | c || c | c | c | c | }
 \cline{2-10}
 \multicolumn{1}{c | }{} & \multicolumn{5}{c || }{$\Lambda$CDM} & \multicolumn{4}{c |}{Extensions} \\ \hline 
 Parameter & $H_0$ & $n_s$ & $\Omega_m$ & $\Omega_b$ & $\sigma_8 (z=0)$ & $w_0$ & $w_a$ & $|{f}_{\rm R0}|$ &  $\Omega_{\rm rc} $ \\  \hline 
 mean ($\mu$) &67.3 & 0.966& 0.316& 0.0494 &  0.766& -1 & 0 & 0 & 0 \\ \hline 
 variance ($\sigma$)  & 0.4 & 0.007& 0.009  & 0.032 & 0.004& 0.097& 0.32& $10^{-5.5}$ & 0.173 \\ \hline
\end{tabular}
\caption{Mean model parameter values ($\mu$) and standard deviations ($\sigma$) used in $\tt{ReACT}$ for generating $\Lambda$CDM, $w$CDM, $f(R)$ and DGP matter power spectra for the training, validation and test data.}
\label{tab:params}
\end{table*}
\begin{table}
\centering
\begin{tabular}{| c || c | c | c | c |  }
\hline  
 $z$ & 0.1 & 0.478 & 0.783 & 1.5 \\ \hline 
 $V(z) [{\rm Gpc}^3/h^3]$ & 0.283 & 3.34 & 6.27 & 10.43 \\ \hline 
 $\bar{n}(z) [{h^3}/{\rm Mpc^3}]$ & 0.0013 & 0.0010 & $8.3 \times 10^{-4}$ & $3.6 \times 10^{-4}$ \\ \hline 
\end{tabular}
\caption{Chosen redshift bins, cosmological volume and number density parameters used to construct the Gaussian errors for each redshift bin in Eq.~\eqref{eq:Euclid_error}.}
\label{tab:errorpars}
\end{table}
\section{Training the Network}
\label{sec:training}
In this section we discuss the procedure of preparing the training, validation and test data, designing the network architecture and the subsequent hyperparameter optimisation.   

\subsection{Generating and preparing matter power spectra} \label{sec:datagen}

We consider three well studied modifications to $\Lambda$CDM: the $f(R)$ gravity model described in Ref.~\cite{Hu:2007nk}, the Dvali-Gabadadze-Porrati (DGP) brane-world model of Ref.~\cite{Dvali:2000hr} and an evolving dark energy model as parameterised in Refs.~\cite{Chevallier:2000qy,Linder:2002et} ($w$CDM). 
We compute dark matter power spectra for these theories utilising the recently developed code {\tt ReACT} \cite{Bose:2020wch} which calculates modified power spectra using the halo-model reaction method developed in Ref.~\cite{Cataneo:2018cic}. %
We sample the parameter space defining each model and pass the values to {\tt ReACT}, which generates power spectra in four redshift bins $z\in\{1.5,0.785,0.478,0.1\}$ and one hundred $k$-bins in the range $0.01 \leq k \leq 2.5 ~ h/{\rm Mpc}$ at equal intervals in log-space, according to that expected from a Euclid-like survey \cite{Laureijs:2011gra,Blanchard:2019oqi}. Details about the choices of the parameter space, redshift and $k$ ranges, are given in App~\ref{app:matpk}. 

In addition to the aforementioned well studied extensions to $\Lambda$CDM we also include an additional class to represent potential ``unknown'' models.  
Such models would imprint various signatures in the power spectrum that would be correlated in both space and time. 
Since a priori we have no way of knowing what these signals are, we produce a dataset of filters with randomly generated features correlated in $k$ and $z$ before applying these to randomly selected spectra from the set of $\Lambda$CDM, $w$CDM, $f(R)$ and DGP model spectra. 
We describe the method to generate this dataset in App.~\ref{app:randpk}. 

For each of the five models considered, $\Lambda$CDM, $w$CDM, $f(R)$, DGP and random, we use 18,475 examples resulting in a total training dataset size of 92,375 examples. 
Every example is a matrix of dimension $100 \times 4$ with each entry given by the value of the power spectrum in the particular $k$ and $z$ bin. 
Of the 92,375 generated power spectra we set aside 15\% for the validation set with the remainder used in training the BNN.
Furthermore we generate a test set composed of 2500 examples per class.
Gaussian noise is then added to each spectrum in accordance to what one would expect from a Euclid-like survey \cite{Feldman:1993ky,Seo:2007ns,Zhao:2013dza,Blanchard:2019oqi}
\begin{equation}
        \sigma_p(k) =  \sqrt{\frac{4 \pi^2}{k^2 \Delta k V(z)} \times \left( P(k) + \frac{1}{\bar{n}(z)} \right)^2 +  \sigma_{sys}^2}  \, .
\label{eq:Euclid_error}
\end{equation}
The redshift dependent survey volume $V(z)$ and the shot noise $\bar{n}(z)$ are presented in Table~\ref{tab:errorpars}. 
In addition, a constant systematic error of $\sigma_{sys}^2 =25 ~{\rm Mpc^6}/h^6$ is included to represent potential modelling inaccuracies. 
This value of $\sigma_{\rm sys}$ is chosen such that we are able to recover the fiducial Planck parameters with $2\sigma$ confidence when performing an MCMC analysis using the nonlinear halofit Planck spectrum as our data vector and Eq.~\eqref{eq:Euclid_error} as our errors, with a $\chi^2$ likelihood.
We leave a thorough analysis of how much this choice affects the results to future work.

For each of the original set of 92,375 examples we generate ten spectra each with a different realisation of the Gaussian noise, ensuring that on top of recognising deviations from particular models, the network is more robust to different noise realisations for the same model. 
In total, the number of training and validation examples is given by 923,750.

Finally, to ensure the data passed to the BNN are of comparable orders of magnitude across all scales and redshift bins we normalise each training example to a reference $\Lambda$CDM power spectrum with a cosmology given by the mean values in Table~\ref{tab:params}. 

\begin{figure*}
\centering
\includegraphics[width=.49\textwidth]{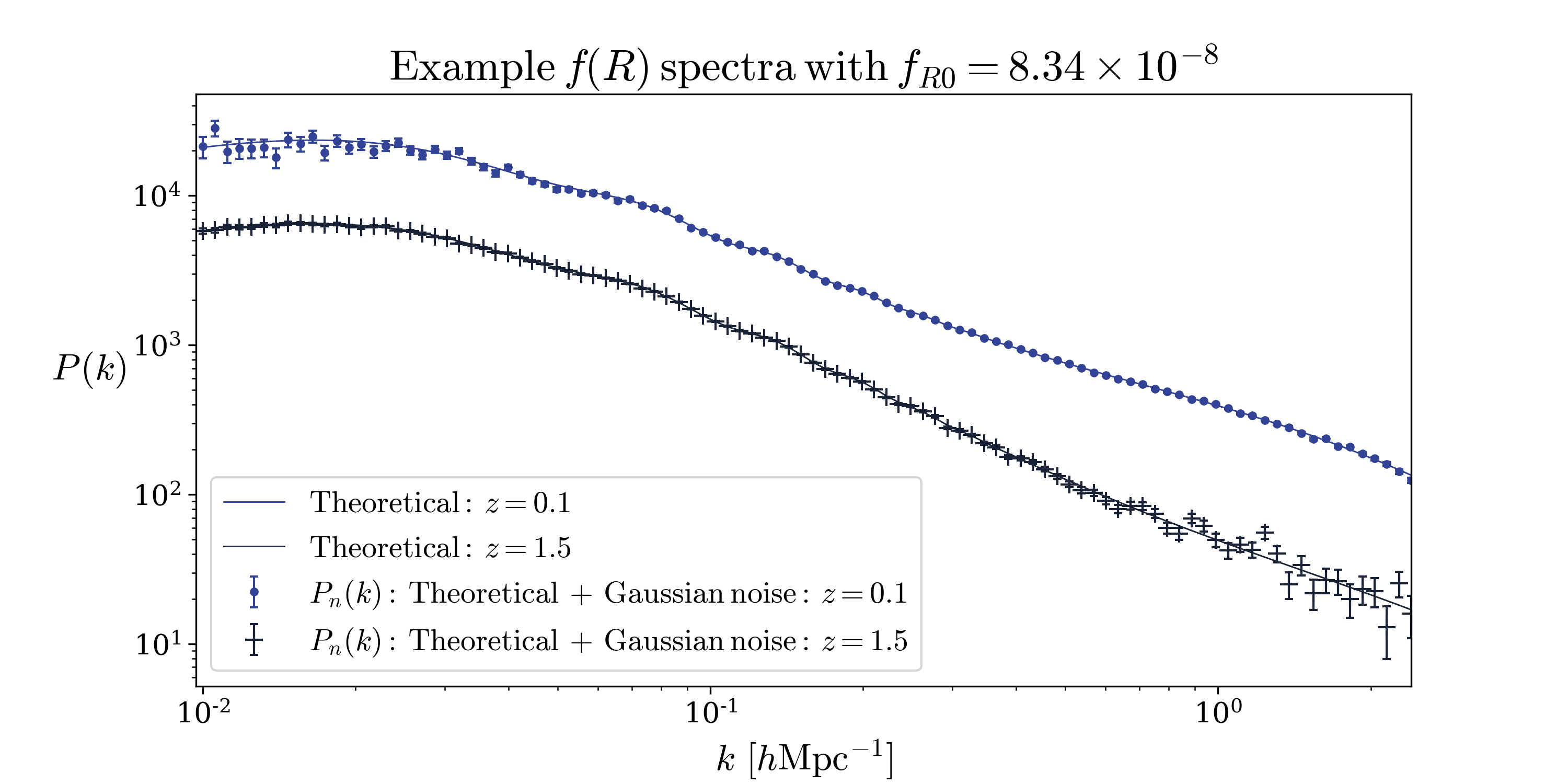}
\includegraphics[width=.49\textwidth]{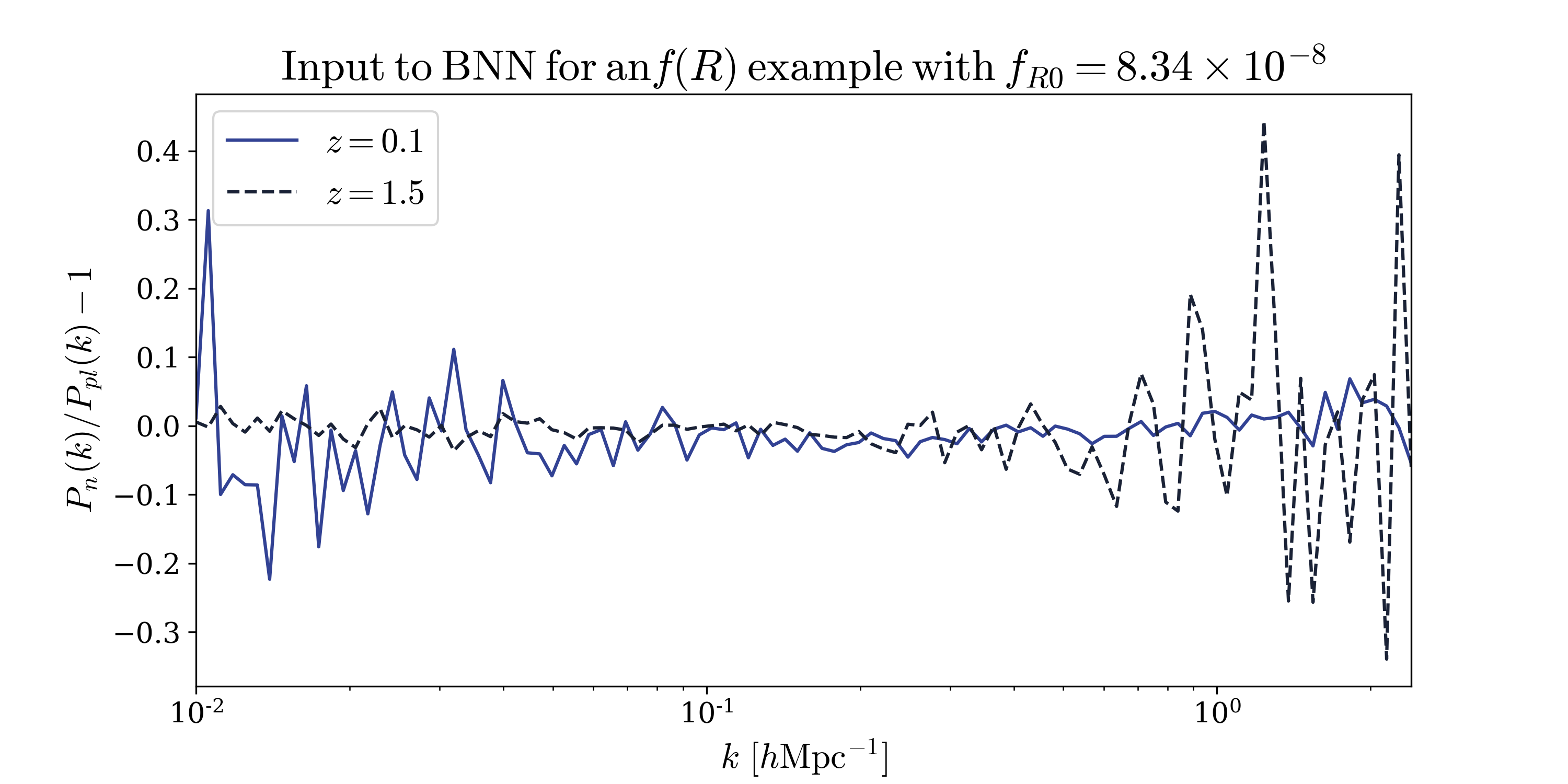}
\caption{\textbf{Left}: A pair of example $f(R)$ spectra with $f_{R0}= 8.34 \times 10^{-8}$ at redshifts $z=1.5$ and $z=0.1$ generated using {\tt ReACT} with the additional Gaussian noise.
Note that at low redshift cosmic variance dominates at low $k$ and at high redshift the shot noise dominates at high $k$.
\textbf{Right}: After normalising the noisy spectrum $P_{n}(k)$ by a fiducial Planck spectrum $P_{pl}(k)$ and centering around zero the spectra are ready to be passed to the BNN.
Due to this normalisation choice the BNN is trained to detect deviations from this fiducial Planck spectrum. 
Note that in practice all four redshift bins are passed to the BNN.
Despite the presence of such a small modification, the five-label BNN classifies this spectrum as $\Lambda$CDM with only 5\% confidence, favouring the presence of a modification (see Fig.~\ref{fig:five-label_net}).
}
\label{fig:Data_comparison}
\end{figure*}

In Fig.~\ref{fig:Data_comparison} we display the process of how spectra generated with {\tt ReACT} are transformed before being passed to the BNN, including the addition of Gaussian noise followed by normalisation by a fiducial Planck spectrum.
Therefore the network is trained to detect deviations from $\Lambda$CDM for different noise realisations and choice of standard cosmological parameters.

\subsection{Training and optimisation}
\label{sec:optimisation}
 \begin{table*}
\centering
\begin{tabular}{| c | c | c | c | c | c |   }
 Operation Layer& Number of filters & Size & Stride  & Output size & Number of parameters \\ \hline \hline 
Input & - & - & - &  100x1x4 & - \\ \hline\hline 
Convolution 1D flipout & 8 & 10 & 2 &  46x1x8 & 648\\ \hline
Max Pooling 1D & - & 2 & 2 &  23x1x8 & -  \\ \hline
Batch normalization & - & - & - &  23x1x8 & 32  \\ \hline\hline 
Convolution 1D flipout & 16 & 5 & 2 &  10x1x16 & 1296\\ \hline
Max Pooling 1D & - & 1 & 2 &  9x1x16 & -  \\ \hline
Batch normalization & - & - & - &  9x1x16 & 64  \\ \hline\hline 
Convolution 1D flipout & 32 & 2 & 1 &  8x1x32 & 2080\\ \hline
Batch normalization & - & - & - &  8x1x32 & 128  \\ \hline\hline 
Global Average Pooling & - & - & - &  32 & -  \\ \hline\hline 
Dense flipout & - & - & - &  32 & 2080\\ \hline
Batch normalization & - & - & - &  32 & 128  \\ \hline\hline 
Dense flipout & - & - & - &  5/2 & 325/130  \\ \hline
\end{tabular}
\caption{Description of the network's architecture}
\label{tab:architecture}
\end{table*}
\begin{figure*}
    \centering
    \resizebox{0.8\textwidth}{!}{
    \includegraphics{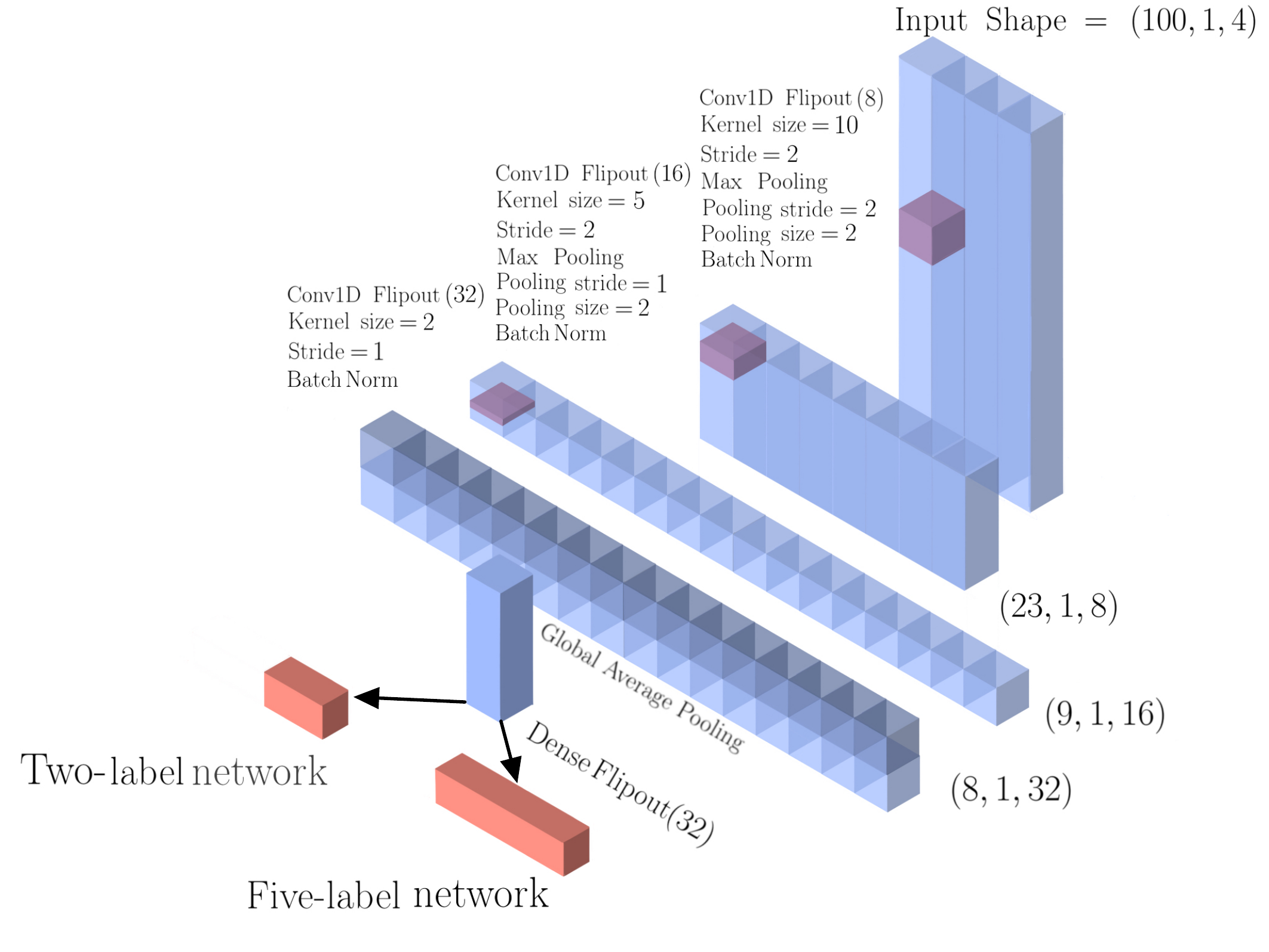}}
    \caption{Depiction of the BNN architecture employed for both the five-label and two-label classification tasks.
    The height of each block illustrates the dimension size for each layer while the number of blocks per layer corresponds to the number of filters. 
    Additionally the dense blocks embedded in the first three transparent layers indicate the kernels for the first three one-dimensional convolutional layers scaled by their respective size.
    }
    \label{fig:model_architecture}
\end{figure*}
\begin{figure*}
    \centering
    \includegraphics[width=.49\textwidth]{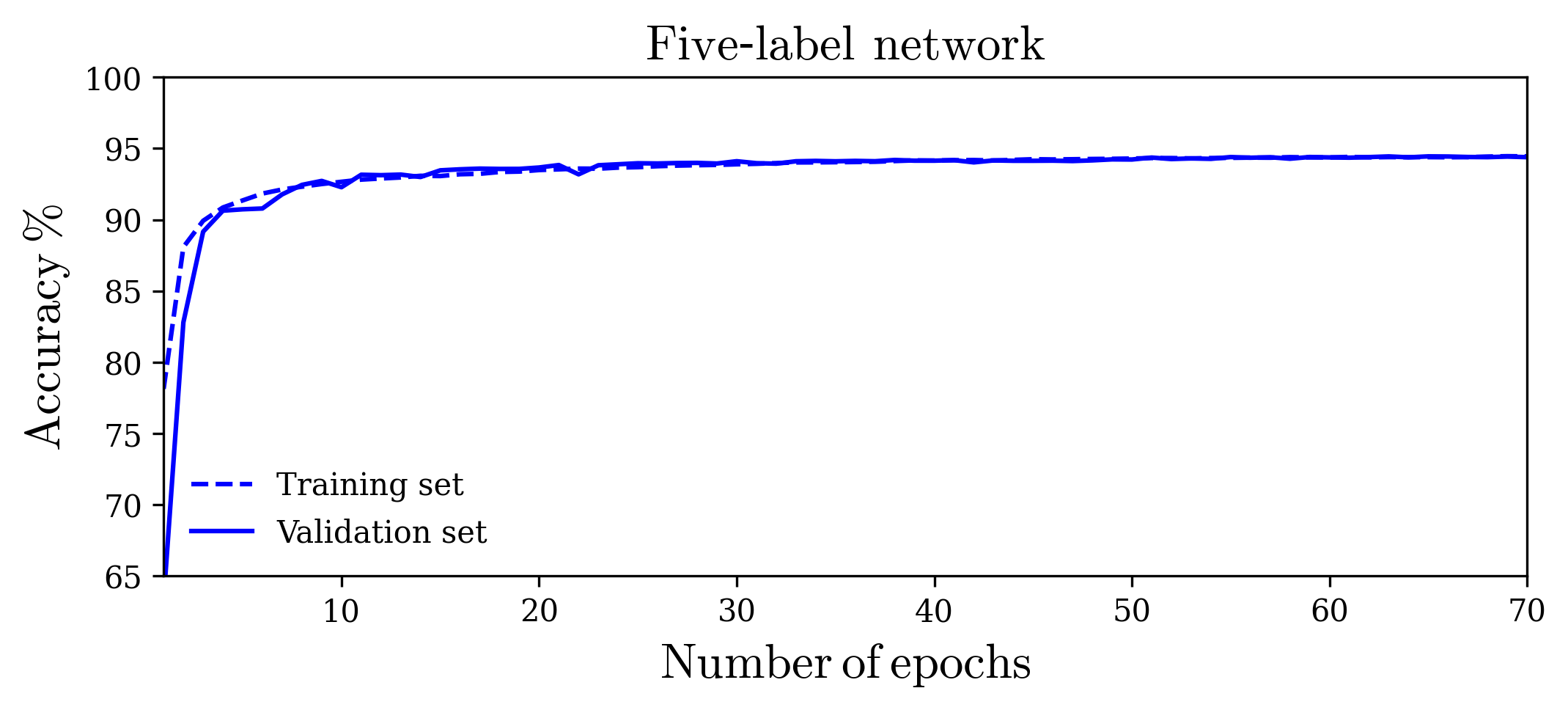}
    \includegraphics[width=.49\textwidth]{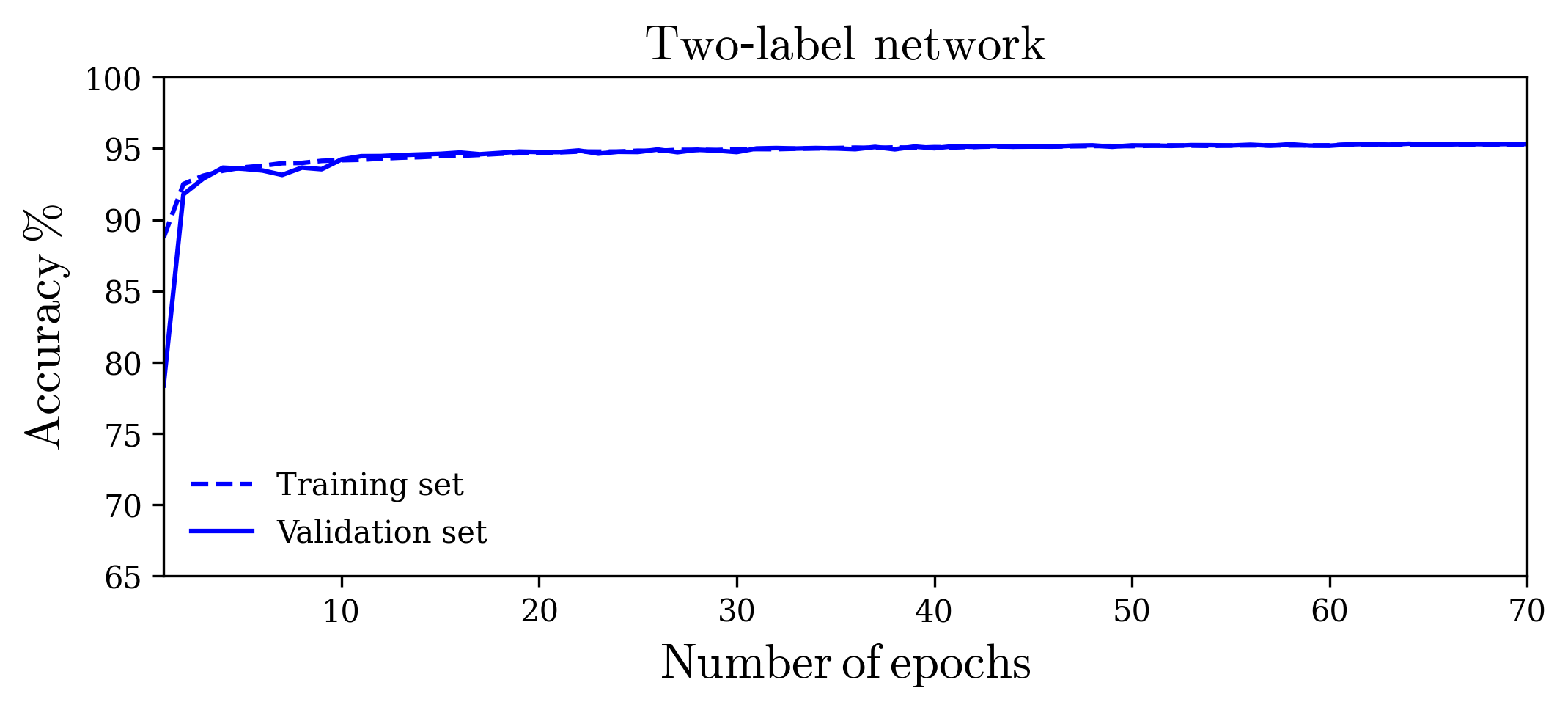}
    \caption{\textbf{Left}: Evolution of the training and validation accuracy for the five-label network. 
    The accuracy on both data sets stabilises at $\sim 94.4 \%$ indicating that the five-label network is relatively robust. 
    \textbf{Right}: Evolution of the training and validation accuracy for a network trained to classify spectra between $\Lambda$CDM or non-$\Lambda$CDM, reaching 95.3\% accuracy for both datasets.
    Note that the accuracy for both networks is evaluated with the output from a single pass through the BNN for all the training and validation spectra and it therefore simply gives an estimate of the network's performance. 
    }
    \label{fig:LC_5labels_1D}
\end{figure*}
In this work we are concerned with the capability BNNs possess in tackling two questions.
The first is how effective BNNs can be in recognising the distinct features in the power spectrum for a particular modification to $\Lambda$CDM, such as $f(R)$ or DGP.
The second is the ability of BNNs to detect a deviation from $\Lambda$CDM in the power spectrum irrespective of the particular modification. 
In practice, we train two BNNs with the same architecture, the first for five labels divided between $\Lambda$CDM and the four extensions and the second trained to distinguish between the two labels $\Lambda$CDM and non-$\Lambda$CDM.
Due to the fact that there are only four redshift bins it is beneficial to treat the data as four separate time-series and use one-dimensional convolutional layers.
Treated this way, the spectra are passed to the network with dimension $100 \times 1 \times 4$, or in analogy with image classification tasks, as $100 \times 1$ pixel images with four channels. 
The architecture of the network used to train both the five-label and two-label network is displayed in Fig.~\ref{fig:model_architecture}.
Initially the structure consists of three 1D convolutional flipout layers with $8, 16$ and $32$ filters, kernel sizes of $10, 5$ and $2$ with strides of $2,2$ and $1$ respectively. 
Each of the first two 1D convolutional layers are followed by a max pooling layer with a pool size of 2 and a pooling stride of 2 for the first max pooling layer and a pooling stride of 1 for the second max pooling layer.    
After both of these max pooling layers there is a batch normalisation layer.
Following the final convolutional layer there is a global average pooling layer to reduce the filter size to one in order to pass it to a dense layer with 32 nodes.
Finally, after a further batch normalisation there is a softmax layer consisting of five or two neurons for either the five or two-label networks respectively.
The network's architecture is summarized in Table \ref{tab:architecture}.
The five-label and two-label networks consist of 6605 and 6410 trainable parameters respectively. 
We set the initial learning rate $\text{lr}_0$ to be 0.01 with a decay rate 0.95 such that with a training set size $M$ and at each epoch $e$ the learning rate is
\begin{equation}
\text{lr}(e) = \text{lr}_0 \times 0.95^{(e/M)} \, .
\end{equation}
The batch size was set to $500 \times 5$. Each batch is composed of an equal number of power spectra for each of the labels in the training data. During training, we adjust the training set size by dropping a random subset of the data in order to have an integer number of batches of the same size. 
In Fig.~\ref{fig:LC_5labels_1D} we show the evolution of the accuracy for a network trained to classify between $\Lambda$CDM, $w$CDM, $f(R)$, DGP and random spectra (top panel) and a network trained to distinguish simply between $\Lambda$CDM from non-$\Lambda$CDM (bottom panel).
In each classification task for both the training and the validation sets the five-label network asymptotes to an accuracy of $94.4\%$ and the two-label network asymtotes to a training accuracy of $95.3\%$.
Bear in mind that the accuracy is evaluated by passing examples once through the BNN with a single draw from the weight distribution and therefore only approximates the BNN's overall performance.
Note also that despite the fact the overall accuracy of the two-label network is slightly greater, it does not necessarily imply that it is generally better at detecting deviations from $\Lambda$CDM in the power spectrum (see Sec.~\ref{sec:results}).

\section{Results}
\label{sec:results}

In this section we determine the ability of both the five and two-label networks to classify previously unseen matter power spectra and perform tests to determine the robustness of the method.
In Sec.~\ref{sec:test_set} we study the overall performance of the network on the test set, followed by a calibration check in Sec.~\ref{sec:calibration} as well as  test the robustness of the five-label BNN against variations in the training set (Sec.~\ref{sec:varTrain}). We then evaluate the performance on individual spectra in Sec.~\ref{sec:example}, including a study of the impact of noise on the classification in Sec.~\ref{sec:noise} and a comparison of the two and five-label networks in Sec.~\ref{sec:fivevtwo}.
We examine the ability of each network to recognise out-of-distribution examples which were not included in the training set in Sec.~\ref{sec:out_of_dist} before studying the constraints each network is capable of placing on the model parameters in Sec.~\ref{sec:constraints}. We finally comment on the relevance for future experiments in Sec.~\ref{sec:experiments}.

\subsection{Performance on the test set}
\label{sec:test_set}

\begin{figure*}
 \centering
    \includegraphics[width=.5\textwidth]{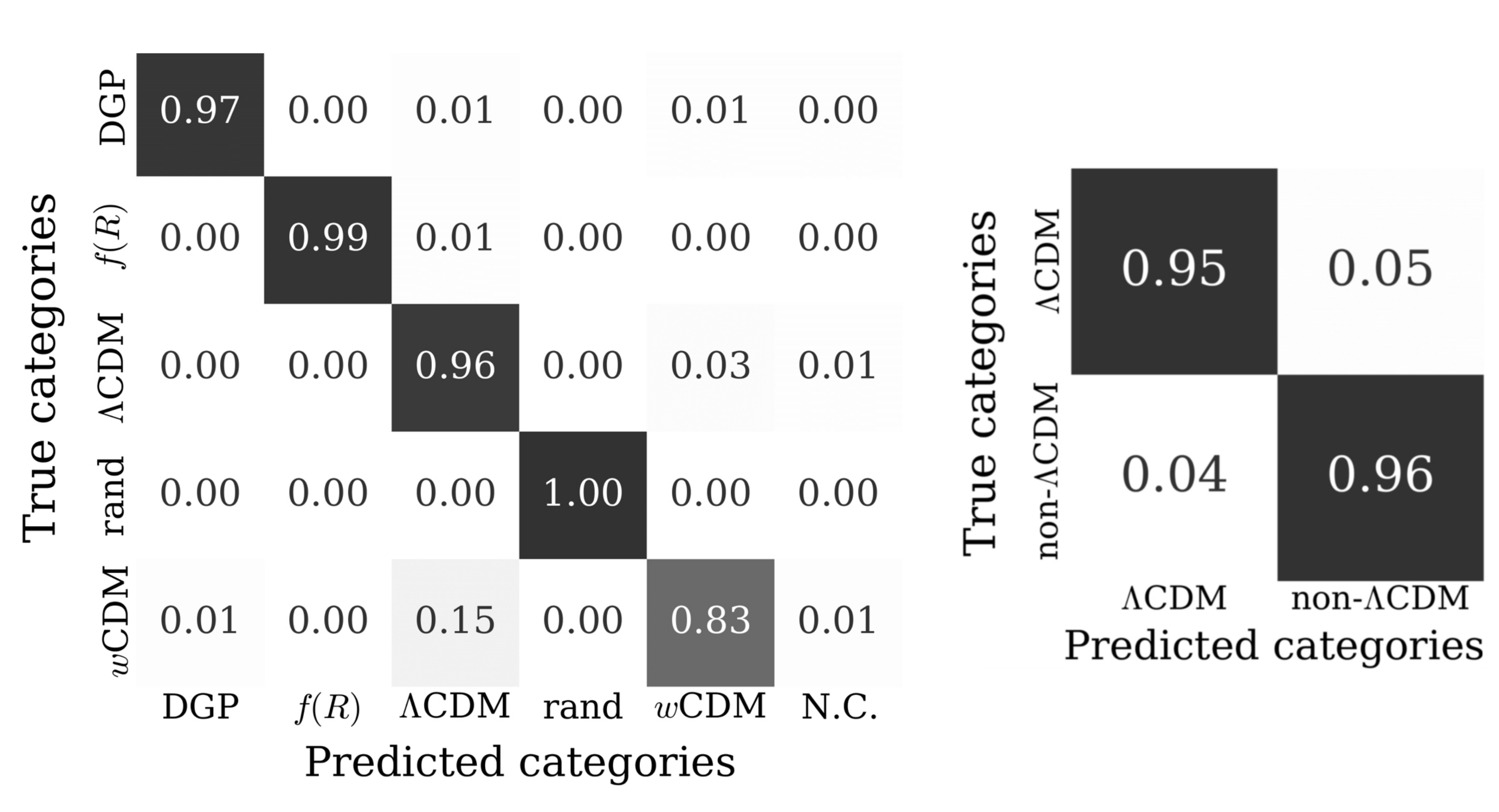}
  \caption[CONVERGENCE]{
  The confusion matrices for the five-label network (Left) and the two-label network (Right) display the percentage of examples per class that are assigned to each class by the network. 
  A classification is obtained by assigning the label to be the maximum multinomial mean $\mu_{i}$ as long as it exceeds the threshold $p_{th}$. 
  If no $\mu_{i}$ exceeds $p_{th}$, the example is considered to be ``Not Classified'' (N.C.). 
  } 
\label{fig:test_results}
\end{figure*}
\begin{figure*}
\centering
    \resizebox{0.3\textwidth}{!}{
    \includegraphics[width=0.3\textwidth]{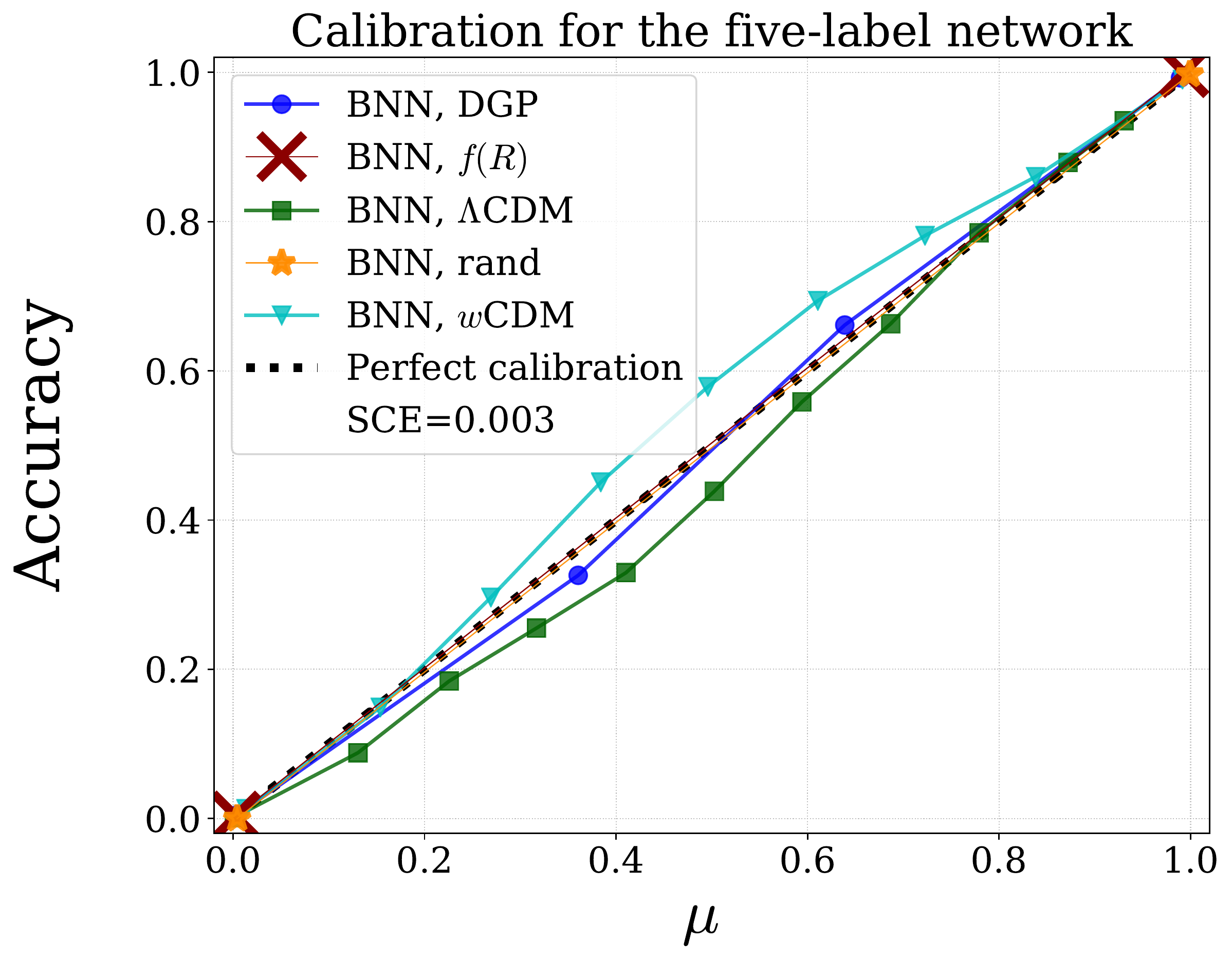}
     }
     \resizebox{0.3\textwidth}{!}{
    \includegraphics[width=0.3\textwidth]{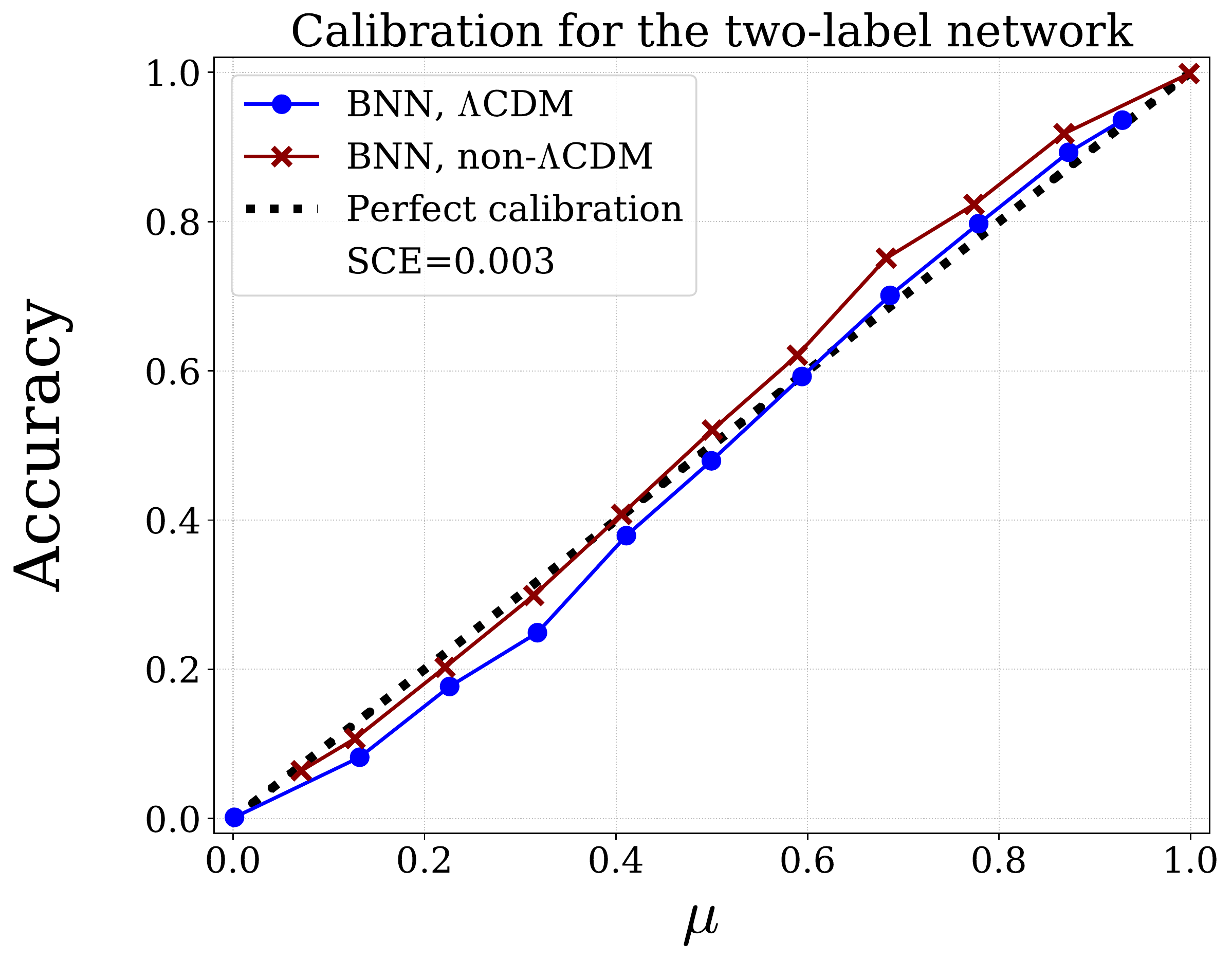}
     }
  \caption[CONVERGENCE]{
  The reliability diagrams for both the five-label (Left) and the two-label (Right) BNNs display the predictions of $\mu$ for examples in the test set divided into bins containing at least $0.5\%$ of the total test examples plotting against the resultant test accuracy evaluated on each bin. 
  A perfectly calibrated network corresponds to the bisector with the deviation from ``perfect calibration'' measured by the SCE defined in Eq.~\ref{SCE} also reported.
  } 
\label{fig:rel_diagram}
\end{figure*}
Now that the network has been trained, the next step is to evaluate its performance on the test set in order to determine how capable it is in classifying previously unseen examples.
To this end, ten copies of every test example are made with different noise realisations added to the same underlying spectra.  
By computing the average of the output after five hundred MC samples using Eq.~\eqref{probabilityBayes}, each example's label is assigned to be the maximum $\mu_{i}$ as long as it exceeds the threshold value $p_{\text{th}}=0.5$.
If no $\mu_{i}$ is greater than the threshold the example is assigned the label ``Not Classifed'' (N.C.).
The resulting overall test accuracies are $94.9 \%$ and  $95.8 \%$ for the five and two-label networks respectively.
As these results are comparable to the training and validation accuracies in Fig.~\ref{fig:LC_5labels_1D} the network can be considered to be robust. 
In Fig.~\ref{fig:test_results} we show the confusion matrices for each network which provide information on the percentage of examples from each class that are classified accurately and if not, what class they were erroneously classified into.
Theories that show a greater degree of degeneracy in their effects on the matter power spectrum are more likely to be classified incorrectly.
For the five-label network, the strongest degeneracy exists between $w$CDM and $\Lambda$CDM, likely because the signatures of $w$CDM occur at length scales where the noise can dominate. 
Indeed, $w$CDM modifications appear at the level of the cosmological background. In contrast, the other theories considered here can affect the higher order perturbations which leave a direct imprint on the power spectrum.
Following $\Lambda$CDM, $w$CDM also possesses a slight degeneracy with DGP. 
By contrast, only 1\% of $f(R)$ examples were mis-classified which correspond to spectra with small values of $f_{R0}$ that are noise-dominated.
The high efficacy the BNN has in detecting $f(R)$ models warrants a more detailed analysis 
which we discuss in Sec.~\ref{sec:specialist}. 

\subsection{Calibration}\label{sec:calibration}

Another important test is to ensure the probabilities given by the network represent the likelihood for a prediction to be correct.
If this holds, the network is said to be well-calibrated.
By definition, a model is perfectly calibrated if the accuracy on all examples classified with probability $p$ is $p\times100 \%$. 
Ensuring that DNNs are well-calibrated is a key step in assessing their reliability \cite{DBLP:conf/icml/GuoPSW17,kuleshov2018accurate,Nixon_2019_CVPR_Workshops}.

In Fig.~\ref{fig:rel_diagram} we present reliability diagrams for both the two and five-label networks which are constructed as follows. 
First, we divide predictions for $\mu$ into bins. 
The number of bins for each class, or component $\mu_{i}$, is chosen such that each bin contains at least $0.5\%$ of the total number of examples in the test set in order to avoid a large variance. 
We then compute the accuracy in every bin for each class.
Let $B_i$ denote the set of bins for the $i$-th class, $n_{bi}$ the number of predictions in bin $b$ for class $i$ and $\text{acc}(b,i)$ and $\hat{\mu}(b,i)$ the corresponding accuracy and average probability for each $b$ and $i$ respectively.
Shown in Fig.~\ref{fig:rel_diagram} are the reliability diagrams displaying how $\text{acc}(b,i)$ varies with $\hat{\mu}(b,i)$ for both the five-label network and the two-label network. 
For the $f(R)$ and random examples in the test set we find the probability is always either very close to 1 or 0 resulting in only two bins.
By construction, the reliability diagram would result in a straight line for a perfectly calibrated network.
We can quantify the deviation from perfect calibration by computing the Static Calibration Error (SCE), defined as \cite{Nixon_2019_CVPR_Workshops}
\begin{equation}\label{SCE}
    \text{SCE} =\frac{1}{N}\sum_{i=1}^{N} \sum_{b=1}^{B_i} \frac{n_{bi}}{N_{\text{tot}}} \left| \text{acc}(b,i)- \hat{\mu}(b, i)\right| \, ,
\end{equation}
where $N_{\text{tot}}$ is the total number of test examples and $N$ is the number of labels. 
From their values displayed in Fig.~\ref{fig:rel_diagram} we find that both networks are well-calibrated with an SCE of $ 0.3\%$. 
Furthermore, we verified this value remains stable under changes in the number of bins and never exceeds $ \lesssim 0.5\%$.

\subsection{Robustness against variations in the training set}\label{sec:varTrain}

In this subsection we evaluate the impact of using the confidence introduced in Sec.~\ref{sec:construction_conf} to detect extensions to $\Lambda$CDM in the matter power spectrum and its usefulness in taking into account uncertainty due to the presence of noise in the training set. 
Recall that, even when marginalised over the weights, the network's output is still conditioned on the training data (see Eq.~\eqref{probabilityBayes}). 
Despite the fact we include multiple realisations of the noise for each clean spectrum to ensure the BNN is more robust to variations in the training set, there is no guarantee this eliminates significant fluctuations in the result with slight variations in the training set.    
However, if the resulting classification of a particular example is highly dependent on the specific realisation of the noise during training, the associated uncertainty will be large for a similar example at test time.
As the covariance matrix in Eq.~\eqref{fullCov} contains an estimate of the aleatoric uncertainty, the classification confidence in Eq.~\eqref{pGauss} should be lower for such noise-dependent examples. 

To explore this issue we train a second five-label network with an alternate partitioning of the data into training and validation sets before evaluating the probability $\mu$ for every example in the test set for both networks.
Note that the same realisation of the noise was added to each test example to ensure any variation in the result cannot be accounted for by the variation in the noise at test time.  
We find that for 243 test examples, or $2\%$ of the test set, each network gives different predictions. 
When considering our estimated confidence however, in 217 of these cases, or $89\%$ of the discrepancies, both networks yield a classification confidence of $<1\sigma$. 
Of the remaining 26 discrepancies, only 3 give inconsistent predictions and in 23 cases one of the two network predictions has a confidence of $<1\sigma$. 
\begin{figure*}
\vspace{0.5cm}
\resizebox{.9\textwidth}{!}{
\includegraphics{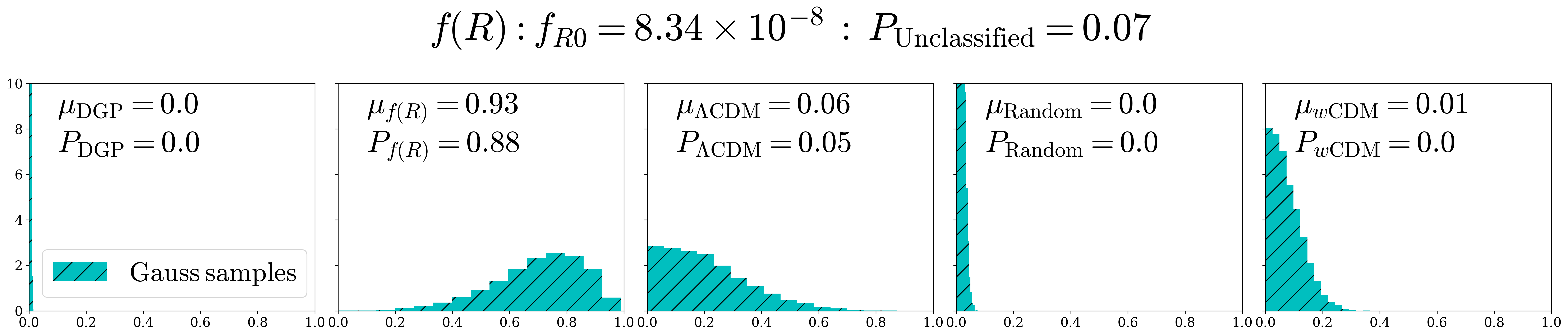}
}
\vspace{0.5cm}
\resizebox{.9\textwidth}{!}{
\includegraphics{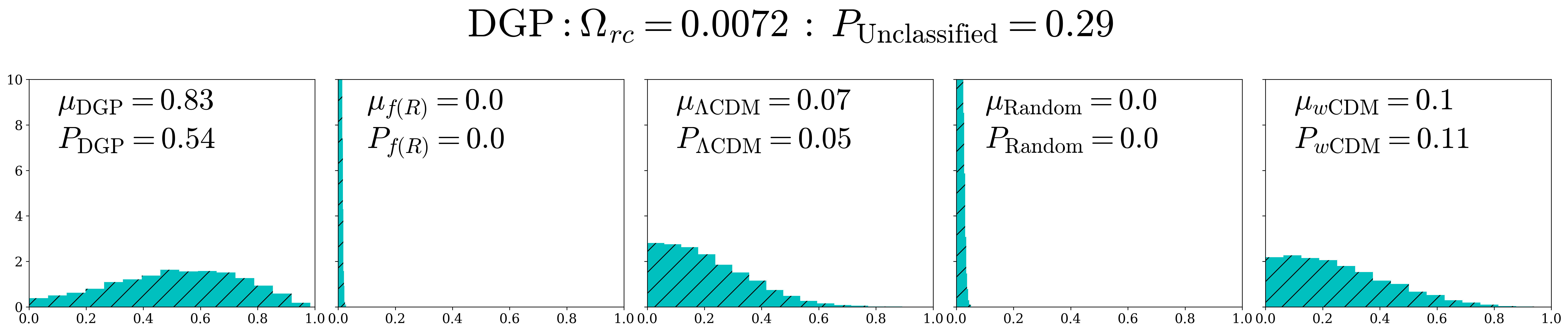}
}
\resizebox{.9\textwidth}{!}{
\includegraphics{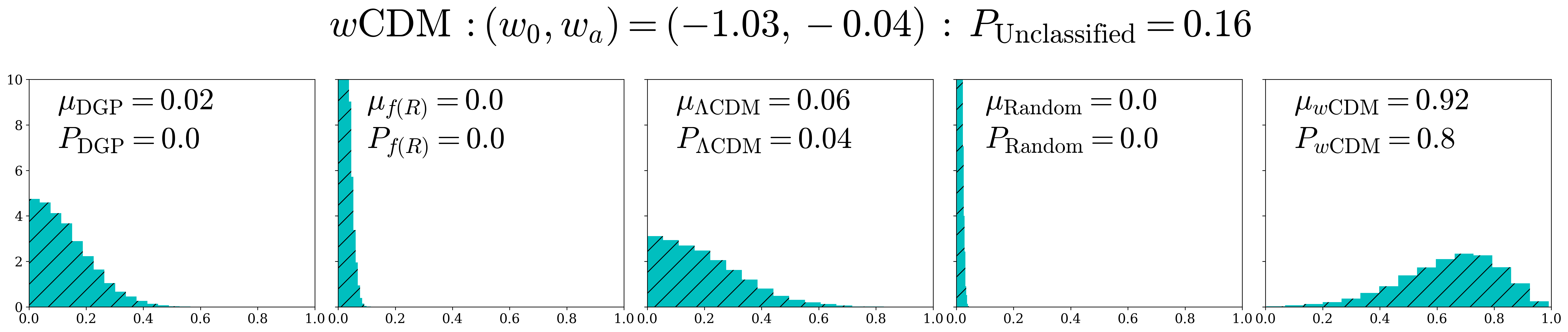}
}
\caption{Examples of a $w$CDM, $f(R)$ and DGP model which the BNN can correctly identify as non-$\Lambda$CDM at  our chosen confidence level. 
We compute $\mu$ and $\Sigma_{q_{\theta}}$ as described in Sec.~\ref{sec:construction_conf}, use them to construct the probability $\mathcal{F}$ defined in Eq.~\eqref{probDist}, then sample from this distribution. The corresponding samples are shown in green. The probabilities denoted by $P_I$ ($I= \text{DGP}, f_R,\Lambda\text{CDM}, \text{Random}, w\text{CDM}$) correspond to the fraction of samples where the $I$-th component of the sample lies above $0.5$, i.e. Eq.~\eqref{pGauss}. If a sample has no component above  $0.5$, it is considered as unclassified. The fraction of samples for which this happens gives $P_{\text{Unclassified}}$.
}
\label{fig:five-label_net}
\end{figure*}
Since each discrepancy involves spectra with very small deviations from $\Lambda$CDM,
we generate an additional dataset of 200 example spectra for each of the three extenions DGP, $f(R)$ and $w$CDM with narrower ranges for the model parameters in the regime where each network may give different predictions, namely $f_{R0} \in \left[3\times10^{-8}, 1\times 10^{-7} \right] $, $\Omega_{rc} \in \left[0.002, 0.06\right]$, $w_{0} \in\left[-1.025, -0.975 \right]$ and $w_{a} \in \left[-0.1,0.1 \right]$. 
In this case, we find a discrepancy in $10\%$ of the dataset but in all these cases at least one network has a confidence of $<1\sigma$.
In $94\%$ of these discrepant examples both networks yield a classification confidence of $<1\sigma$ while only in a single case does one network incorrectly classify an example with a confidence $>1\sigma$.   
This analysis suggests that the confidence in Eq.~\eqref{pGauss} is a more realistic indicator of a prediction's reliability with respect to $\mu$.

\subsection{ Illustration on explicit examples}\label{sec:example}

In this subsection we illustrate the new method to compute the confidence, as introduced in \ref{sec:construction_conf}, on explicit examples.

We choose three underlying noiseless spectra belonging to the $f(R)$, DGP and $w$CDM classes, and add a fixed realisation of the noise to each of them, thus mimicking an actual observational situation where the network is given some noisy spectrum to classify.
We choose the parameters and the noise realisation so that the probability of being non-$\Lambda$CDM is around $95\%$ for each example. We will investigate the role of the noise and dependence on the strength of the modifications more extensively in Sec.~\ref{sec:noise} and Sec.~\ref{sec:constraints} respectively.
Following the procedure outlined in Sec.~\ref{sec:construction_conf}, we compute $\mu$ and $\Sigma_{q_{\theta}}$ with the five-label network for each spectrum. These are used to construct the distribution $\mathcal{F}$ using Eq.~\eqref{probDist}. This represents the distribution of possible outcomes of the network, taking into account the epistemic and aleatoric uncertainties. Then, according to the algorithm described in App.~\ref{app:prob} to compute the probabilities in Eq.~\eqref{pGauss}, samples are drawn from $\mathcal{F}$, each sample being a vector of dimensions equal to the number of classes (5) with values between 0 and 1, and where the dimensions $\{0,..., 4\}$ correspond respectively to the classes DGP, $f(R)$, $\Lambda$CDM, Random, and $w$CDM. The fraction of samples where the $I$-th component (with $I\in \{0,..., 4\}$) lies above $p_{th}=0.5$ is determined $\forall I$, which gives the integral in Eq.~\eqref{pGauss}. If a sample has no component above $p_{th}=0.5$, it is considered as unclassified. The fraction of samples for which this happens gives $P_{\text{Unclassified}}$.

In Fig.~\ref{fig:five-label_net} we display the results.
Samples from $\mathcal{F}$ are shown in green. 

The first spectrum we consider (top panel) is an $f(R)$ spectrum with $f_{R0} = 8.34\times 10^{-8}$.
We find it is correctly classified as $f(R)$ with a probability of 88\%, with the remaining probability falling into $\Lambda$CDM with 5\% and unclassified with 7\%. 
This remains consistent with the evaluation of the confusion matrix on the test set which showed there were no $f(R)$ spectra classified as DGP or $w$CDM. 
In contrast, a DGP spectrum with $\Omega_{rc} = 0.0072$ (mid panel) is classified as DGP with only 54\% probability with 11\% $w$CDM, 30\% unclassified and 4\% $\Lambda$CDM showing the stronger degeneracy between DGP and $w$CDM. 
Classifying $w$CDM spectra with small deviations in $w_{0}$ and $w_{a}$ is particularly difficult for the BNN due to their high degree of degeneracy with $\Lambda$CDM and the fact its features appear in noise-dominated regions of the power spectrum.
In this case, a spectrum with a deviation of $(w_{0}, w_{a}) = (-1.03, -0.04)$ (lower panel) is classified as $w$CDM with 78\% probability, 17\% unclassified and 4\% $\Lambda$CDM. 
These modifications represent the minimum deviations from $\Lambda$CDM in each of our chosen extensions before the modifications become noise-dominated and the five-label network determines the spectra to either be unclassifiable or $\Lambda$CDM.  

\begin{figure*}
\centering
\resizebox{.6\textwidth}{!}{
\includegraphics{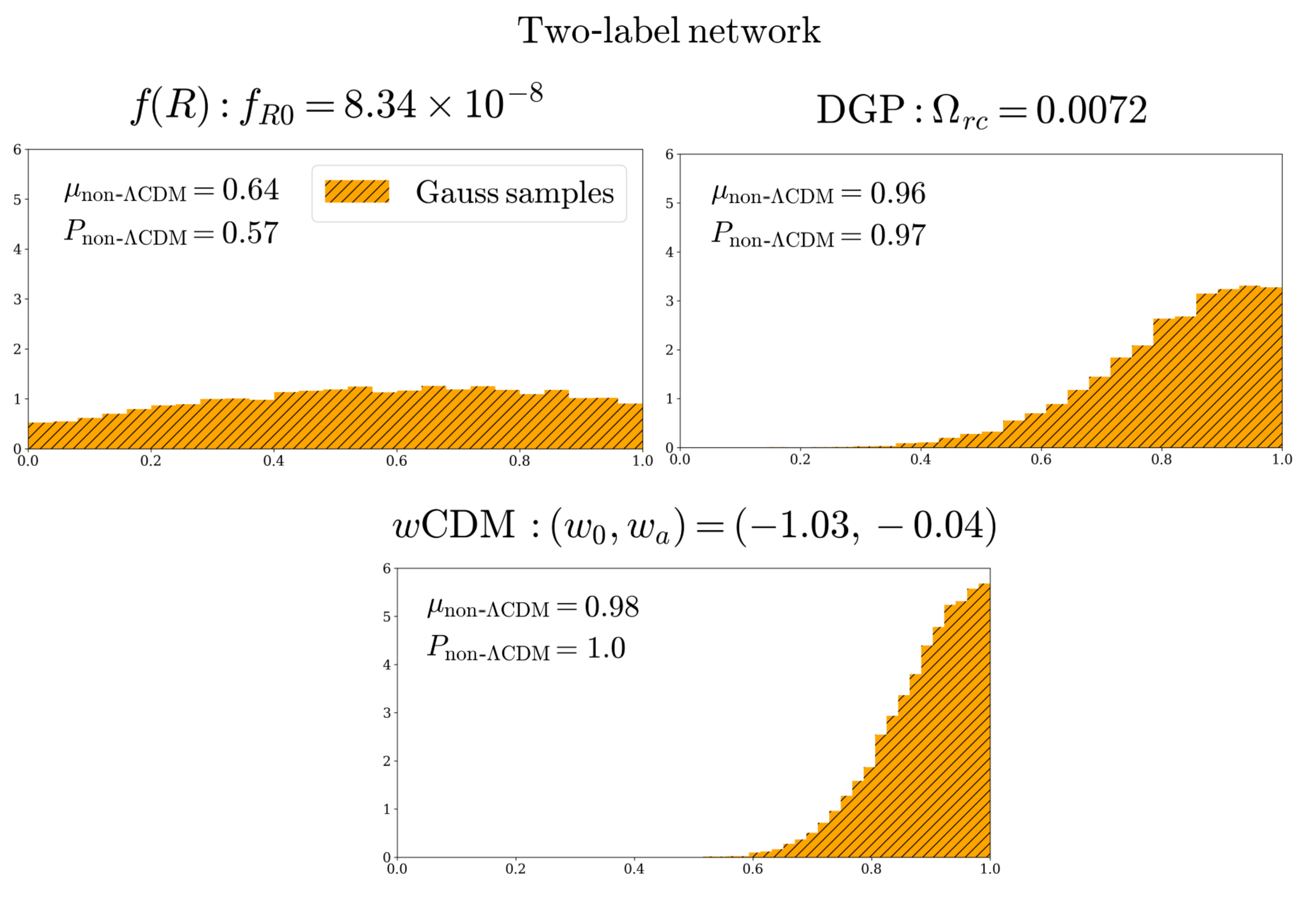}
}
\caption{We display the distributions constructed from the output of the two-label network for the same spectra and Gaussian noise which were passed to the five-label network in Fig.~\ref{fig:five-label_net} in order to compare their relative performance in detecting deviations from $\Lambda$CDM in marginal examples.
Both the $w$CDM and DGP examples are classified as non-$\Lambda$CDM with a higher confidence than the five-label network, however the $f(R)$ spectrum is not correctly classified at high confidence}
\label{fig:two-label_net}
\end{figure*}

We then repeat the procedure for the same noisy spectra with the two-label network.
The result is shown in Fig.~\ref{fig:two-label_net}. 
Note that in the case of two labels the probability of ``Not Classified'' is always zero as it is not possible to have two samples which are simultaneously above $0.5$.  
We find that for DGP and $w$CDM the two-label network classifies the examples correctly with a higher probability than the five-label network.
However the $f(R)$ spectrum is not correctly classified with a high probability.

\subsection{Dependence on the noise}\label{sec:noise}
\begin{figure*}
\vspace{0.5cm}
\resizebox{.9\textwidth}{!}{
\includegraphics{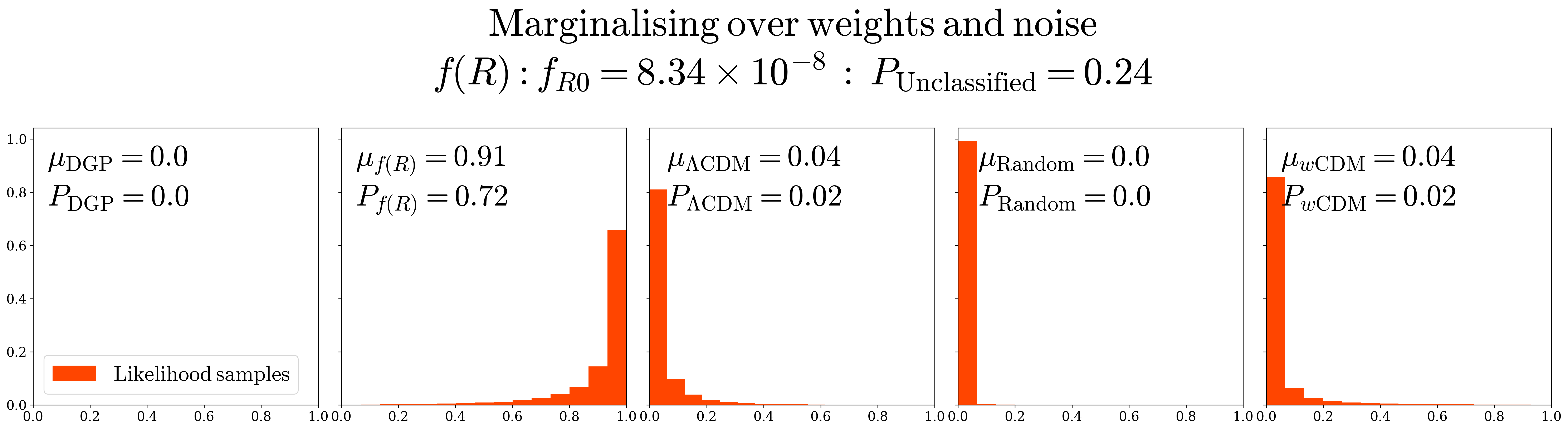}
}
\vspace{0.5cm}
\resizebox{.9\textwidth}{!}{
\includegraphics{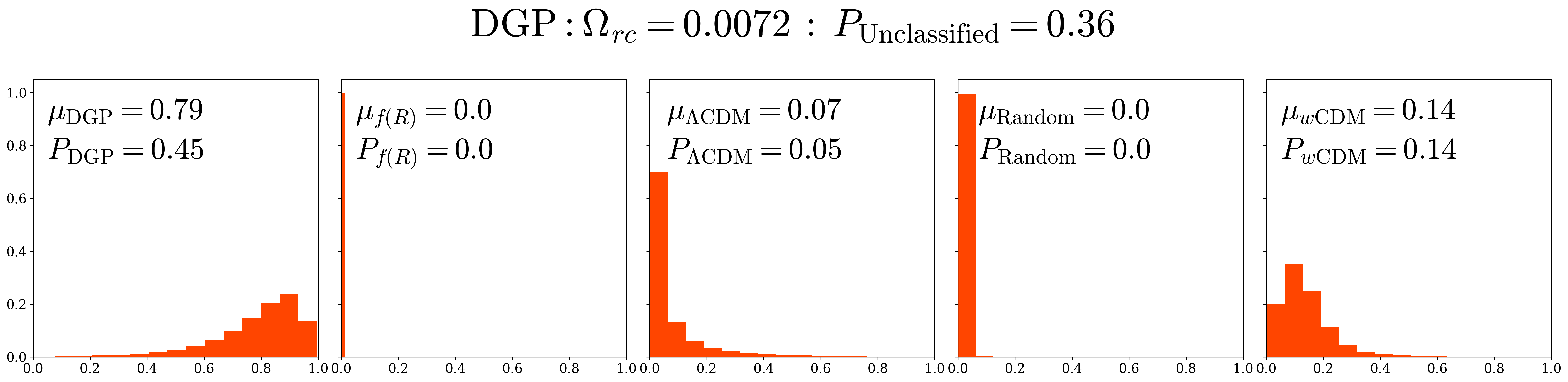}
}
\resizebox{.9\textwidth}{!}{
\includegraphics{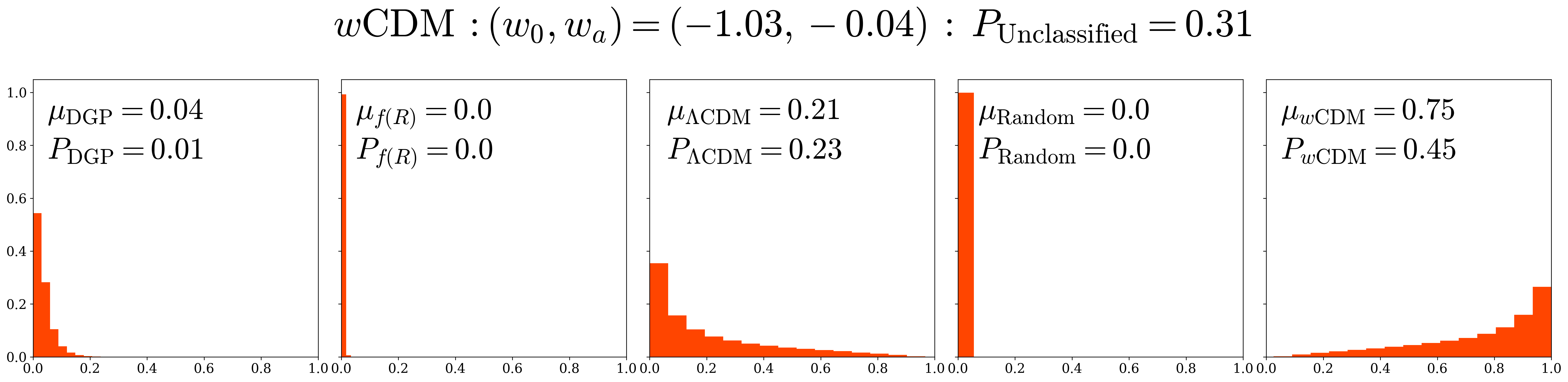}
}
\caption{Taking the same clean $w$CDM, $f(R)$ and DGP spectra as were passed through the network in Fig.~\ref{fig:five-label_net}, we now pass them through the network one thousand times with each pass drawing a new sample from the weight distribution and with a new realisation of Gaussian noise added to the clean spectra.
The resulting distributions therefore give a measure on how the network's output varies when marginalising over the observational noise and the weights.  
}
\label{fig:noise_marginalisation}
\end{figure*}

It is important to emphasise the role noise plays in determining the eventual classification probability for each example in Fig.~\ref{fig:five-label_net}.
In particular, it is possible that a different draw from the Gaussian noise in Eq.~\eqref{eq:Euclid_error} on top of the same underlying clean spectrum could change the resulting classification.
In order to obtain a measure on how much the noise affects the resulting classification for a given underlying spectrum, we compute $\mu$ and $\Sigma_{q_{\theta}}$ again starting from the same underlying noiseless $f(R)$, DGP, and $w$CDM spectra used for Fig.~\ref{fig:five-label_net}, but this time we further average the result over \emph{different} noise realisations for each spectrum. 

In Fig.~\ref{fig:noise_marginalisation} we display the distribution of outputs from the five-label BNN varying the noise realisation. We stress that, differently from Fig.~\ref{fig:five-label_net}, the histograms in Fig.~\ref{fig:noise_marginalisation} do not represent samples from the distribution $\mathcal{F}$ in Eq.~\eqref{probDist}, but are different realisations of $\mu_i$ defined in Eq.~\ref{probabilityBayes} corresponding to different noise realisations on top of a given clean spectrum. This is an illustration of the potential variability of the network's output with noise. 

From the $\mu$ and $\Sigma_{q_{\theta}}$ obtained from averaging over the noise we compute $P_I$, which now becomes a noise-averaged classification probability. The corresponding values are shown in Fig.~\ref{fig:noise_marginalisation}.
This probability gives a measure on how likely it is the network will pick up a deviation from $\Lambda$CDM given the distribution of Gaussian noise in Eq.~\eqref{eq:Euclid_error}.
For example, the $f(R)$ spectrum with $f_{R0} =8.34\times 10^{-8}$ possesses an average detection probability of $\sim 72\%$.
In contrast, the DGP and $w$CDM examples both have noise-averaged detection probabilities of less than $50\%$. 
This implies that, even though the BNN classified each individual example correctly in Fig.~\ref{fig:five-label_net}, for our chosen model parameter values a correct classification was more likely to occur for the $f(R)$ example than for the DGP and $w$CDM examples given another realisation of Gaussian noise. 
Note that this noise-averaged detection probability can be considered to be an invariant measure of the network's performance in classifying spectra with particular values of the model parameters.

\subsection{Performance of five and two-label BNNs}\label{sec:fivevtwo}

To compare the performance of the two-label and five-label networks more robustly, we compute the $\Lambda$CDM classification probability on the test set in each network, as well as compare the $P_{\mathrm{non-}\Lambda \mathrm{CDM}}$ from the two-label network with $1-P_{\Lambda \mathrm{CDM}}$ of the five-label network.
We find that in $\sim98\%$ of the cases where the example is correctly predicted as $\mathrm{non-}\Lambda \mathrm{CDM}$, the five-label network can correctly classify spectra at a higher confidence than the two-label network. 
This is likely a result of the fact that the five-label network, possessing more final classes, can tune its layers to pinpoint specific features of each subclass, resulting in a higher confidence. By contrast, the two-label network needs to compress the information from any deviation into a single class, which can result in lower confidence due to contamination from the classes that are more difficult to distinguish from $\Lambda$CDM.

Of the $2\%$ of spectra where the two-label network was more confident, the probability in the five-label network was either split principally between two non-$\Lambda$CDM classes, not classified or belonged to $w$CDM.   
This indicates that the two-label network may classify non-$\Lambda$CDM spectra which do not belong to any of the classes in the training set more confidently.
Such spectra are more evenly split by the five-label network between separate classes or classified as random (see Sec.~\ref{sec:out_of_dist}).
However, further investigation is required to determine the necessary conditions for the two-label network to outperform the five-label network and vice versa.

\begin{figure*}
\vspace{0.5cm}
\resizebox{.9\textwidth}{!}{
\includegraphics{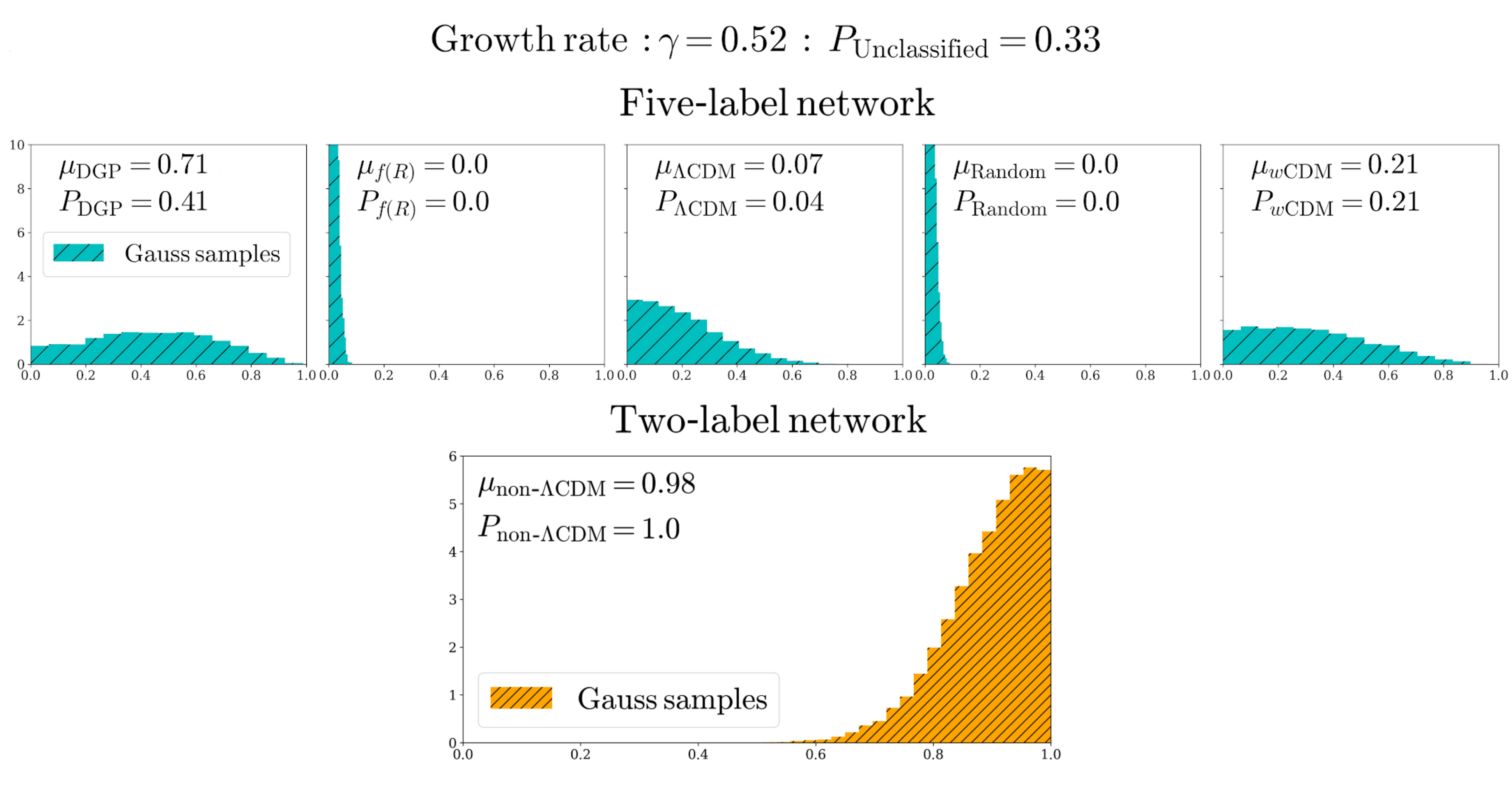}
}
\caption{If an example spectrum generated with a growth-rate parameter of $\gamma=0.52$ with a fixed noise realisation is passed through the five-label (Top) BNN, this estimates this spectrum is not $\Lambda$CDM at the $2\sigma$ confidence level.
However, the remaining probability is distributed between the other labels with no overall favoured class, highlighting the utility BNNs possess in determining a spectrum does not belong to any of the classes in the training set.
If the same spectrum with the same noise realisation is passed to the two-label network (Bottom) it is classified as non-$\Lambda$CDM with a higher confidence than the five-label network.
}
\label{fig:gamma_5label}
\end{figure*}
\begin{figure*}
    \centering
    \resizebox{.9\textwidth}{!}{
\includegraphics{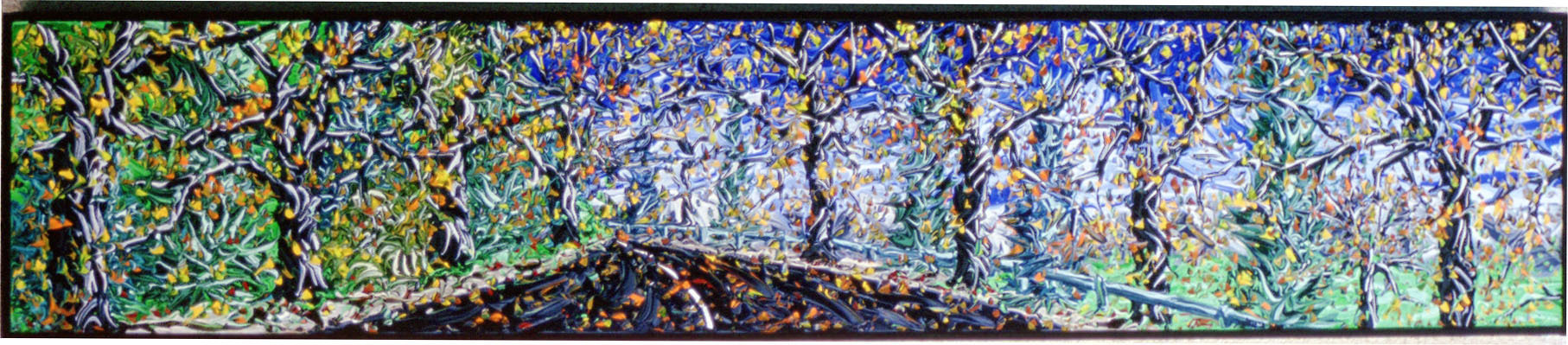}
}
\resizebox{.5\textwidth}{!}{
\includegraphics{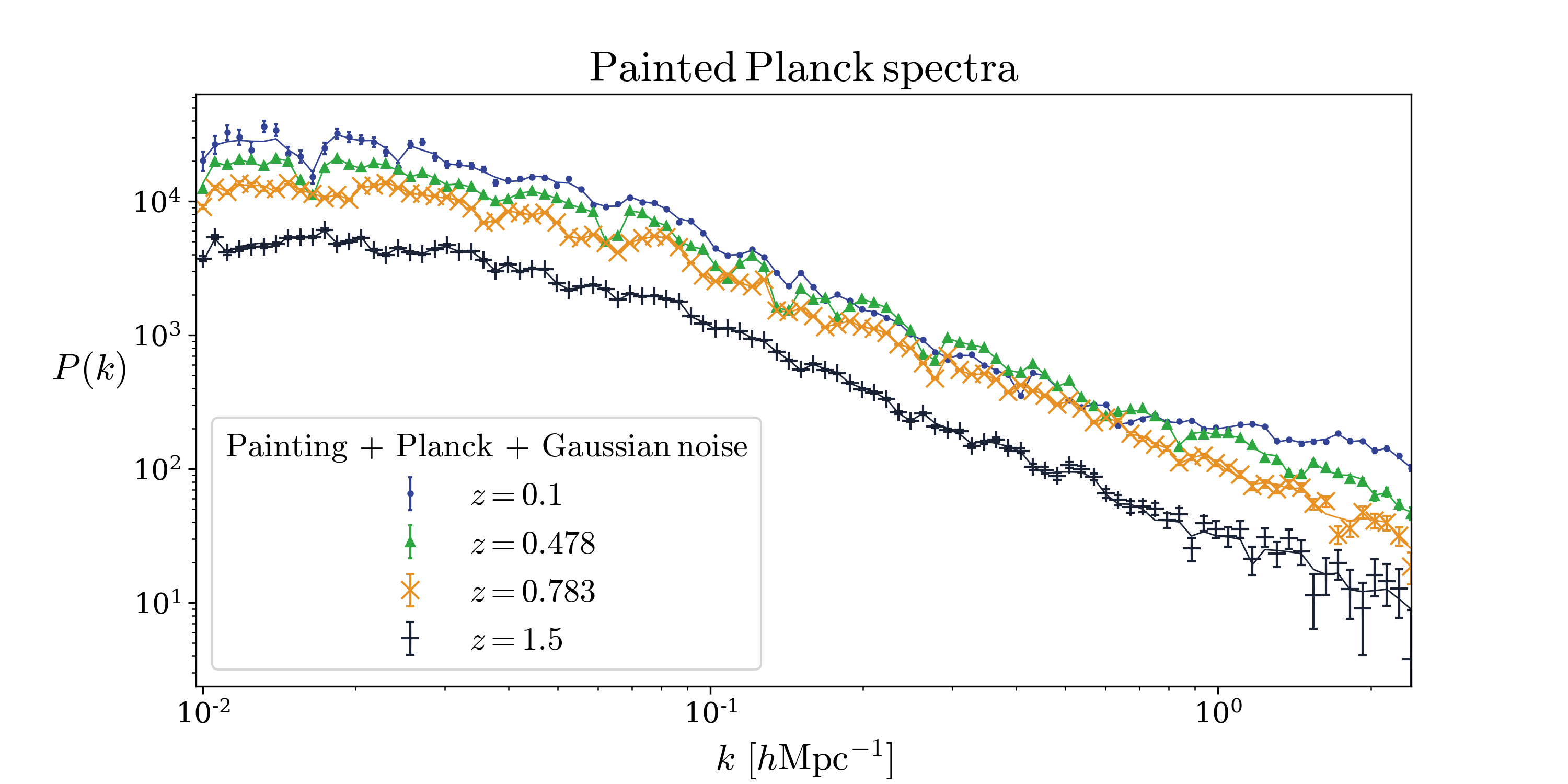}
}
\caption{If an example that does not belong to any of the pre-trained classes is passed to the BNN, in this case the painting ``\emph{Blustery mountain road on an Autumn day}'' by Gregory Horndeski (Top), we find that both the five-label and two-label network classifies it as non-$\Lambda$CDM with a confidence $\gg$ 2$\sigma$. 
In addition, the five-label BNN classifies it as random with extremely high confidence.  %
To pass the painting to the BNN, the pixels were re-binned into a $100\times 4$ pixel grey-scale image which was multiplied by the fiducial Planck spectrum. 
Gaussian noise was then added and the resulting noisy spectra were normalised to the Planck spectrum at each redshift bin (Bottom).
The final normalised spectrum has significant deviations from zero, consistent with the fact that the paining imprints large random deviations from the $\Lambda$CDM spectrum.
}
\label{fig:Horndeski_painting}
\end{figure*}

\subsection{Classification of out-of-distribution examples}
\label{sec:out_of_dist}

To investigate how each BNN classifies examples that do not belong to either the training, validation or test distributions, known as out-of-distribution examples, in this section we examine how each BNN classifies spectra generated both from the growth-index parameter $\gamma$ \cite{PeeblesBook, Wang:1998gt, Linder:2005in} and from a painting by Gregory Horndeski.

The growth-index is a frequently used phenomenological parameterisation designed to pick up deviations in the growth rate of structure from its $\Lambda$CDM value of $0.55$ arising from extensions to $\Lambda$CDM.
The parameterisation is defined by $D'(a) = \Omega_m(a)^{\gamma}$, where $D(a)$ is the linear density perturbation growth factor, $\Omega_m$ is the cosmological total matter density fraction and the prime denotes a logarithmic scale factor derivative. 
To generate nonlinear spectra with varying values of $\gamma$ we first modify the linear power spectrum by applying the following parameterised growth factor
\begin{equation}
D(\gamma;a_f) = \int_{a_i}^{a_f} \left[ \frac{\Omega_{m,0}}{H(a)^2 a^3} \right]^\gamma \frac{a_i}{a} da \, ,
\end{equation}
where $a$ is the scale factor, $H(a)$ is the $\Lambda$CDM Hubble rate and $a_i=0.0001$ is the initial scale factor. 
The modified linear spectrum is then simply $P_L(k,\gamma;a) = D(\gamma;a)^2 P_0(k)$, where $P_0(k)$ is the primordial power spectrum.
The modified nonlinear spectrum is produced by supplying the modified linear spectrum to the same halofit formula \cite{Takahashi:2012em} used in producing the training data. 

We find that, while a spectrum generated with a growth-index of $\gamma = 0.55$ is correctly classified as $\Lambda$CDM, the associated confidence lies between $1\sigma$ and $2\sigma$ reflecting the fact that this parameterisation is only an approximation of $\Lambda$CDM. 
Passing spectra generated with $\gamma = 0.54$ or $\gamma = 0.56$ to the five-label BNN shifts the $\Lambda$CDM classification probability to below $0.5$.
In Fig.~\ref{fig:gamma_5label} we display the sampled classification probabilities and the multinomial mean for each class for a spectrum generated with $\gamma = 0.52$.
With the classification probability of $\Lambda$CDM being $\approx 5\%$, this value of $\gamma $ represents the smallest deviation from $0.55$ such that the network can confidently classify the spectrum as not being $\Lambda$CDM.
Nonetheless, as no probability exceeds $0.5$, no class is favoured.

This ability to determine that a spectrum does not belong to the training set distribution demonstrates a unique capability of BNNs.
Note also that the failure of the five-label BNN to classify a spectrum generated from the growth-index as either $w$CDM, DGP or $f(R)$ further highlights the limitations of the growth-index parameterisation.  
Taking the same spectrum with $\gamma=0.52$ with the same noise and passing it through the two-label network, we find that it is classified as non-$\Lambda$CDM with a higher confidence than the five-label network. 
Although this suggests the two-label network is better suited to placing constraints on the growth-index, given the five-label network did not confidently classify the spectrum into any of the five-labels, it is an open question how useful such constraints would be in constraining more physically motivated models. 
As a further test that the five-label network can identify spectra that do not belong to any known class of physical models, we pass the painting ``\emph{Blustery mountain road on an Autumn day}'' by Gregory Horndeski (see Fig.~\ref{fig:Horndeski_painting}) to the BNN. 
Firstly, we convert it into a grey-scale image with $100\times4$ pixels. 
This then acts as a similar filter to those constructed in App.~\ref{app:randpk} which we then apply to the fiducial Planck spectrum before finally adding Gaussian noise (see Eq.~\eqref{eq:filtertopk}). 
The resulting matrix is then equivalent to a normalised input for the network. 
These deviations are large enough such that the network can accurately determine that the painting is not a $\Lambda$CDM power spectrum.  
However, it is also not ``Not Classifed''. 
Rather, it is classified into the random class with 100\% probability, indicating that the random class is capable of picking up examples that contain deviations which are not comparable to any model included in the training set. 

\subsection{Dependence on the strength of the modification }
\label{sec:constraints}

We have seen in Sec.~\ref{sec:noise} that a more reliable estimator of a BNN's ability to classify a non-$\Lambda$CDM spectrum with a particular modification strength is to pass the spectrum through the BNN multiple times with different realisations of the noise. 
The resultant probability distribution quantifies not only whether a detection is possible, but also how probable it is the noise will alter the classification.
In this section we repeat this procedure for multiple $f(R)$, DGP and $w$CDM power spectra in the parameter range defined by the region where the five-label network transitions from classifying spectra as non-$\Lambda$CDM at low confidence to high confidence.   
Specifically, we build a batch of power spectra composed of different noise realisations on top of the same underlying spectrum and predict the average classification likelihood $\mu$ marginalized over the weights for all the elements in the batch. 
By further averaging the result over the batch, we obtain a noise-averaged classification likelihood for every example. 
Using this to construct the probability distribution in Eq.~\eqref{probDist}, we then compute the corresponding average classification probability that the spectrum belongs to its true class and the average probability it is non-$\Lambda$CDM, defined as $1-P_{\Lambda \mathrm{CDM}}$ for both the five and two-label networks.
This process is then repeated for spectra with different modification strengths.

In order to obtain an estimate on how much the noise can shift the classification for particular values of the model parameters, we also construct a confidence band around the average classification probability as follows. 
First, we remove the noise realisations such that any of the components of its predicted $\mu$ fall below the corresponding fifth or above the corresponding ninety-fifth percentile of the batch.
For each network the upper bound is then obtained by selecting the noise realisation among those remaining such that the probability $\mu_{i}$ (with $i$ being $f(R)$, DGP or $w$CDM for the five-label network, and non-$\Lambda$CDM for the two-label network) is maximised.
The minimum bound is obtained by taking the noise realisation that maximises the difference $ \mu_{\Lambda \mathrm{CDM}}-\mu_{i}$. 
While for the two-label network this is equivalent to minimising $\mu_{ \mathrm{non}-\Lambda \mathrm{CDM}}$, in the case of the five-label network it ensures that the lower bound is the minimum of both $P_i$ and $P_{\mathrm{non}-\Lambda \mathrm{CDM}}$. 
This would not be guaranteed by only taking the noise realisation that minimises $\mu_{i}$, due to the fact that we allow for an unclassified probability.
\begin{figure*}
    \centering
    \resizebox{0.49\textwidth}{!}{
    \includegraphics{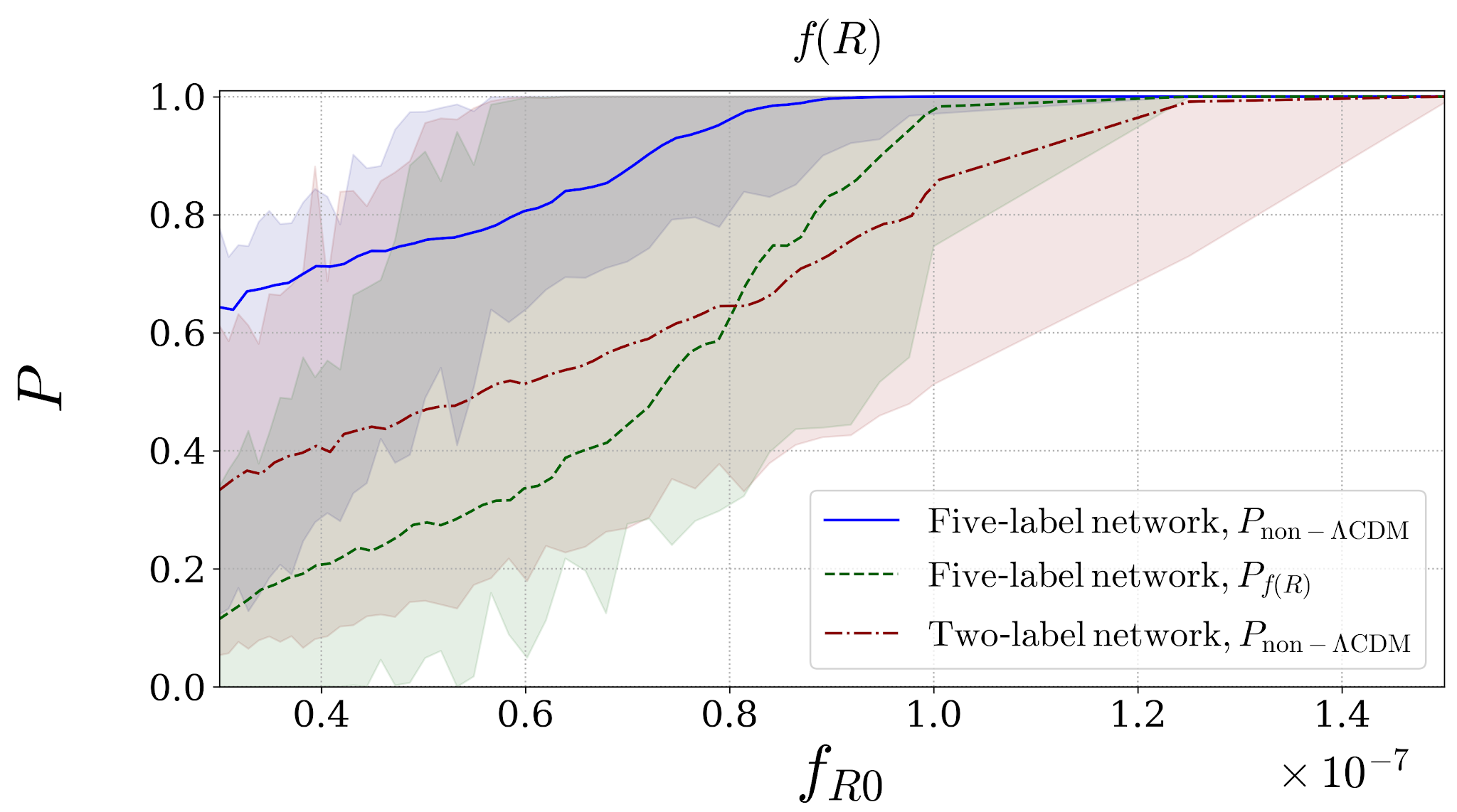}
     }
     \resizebox{0.49\textwidth}{!}{
    \includegraphics{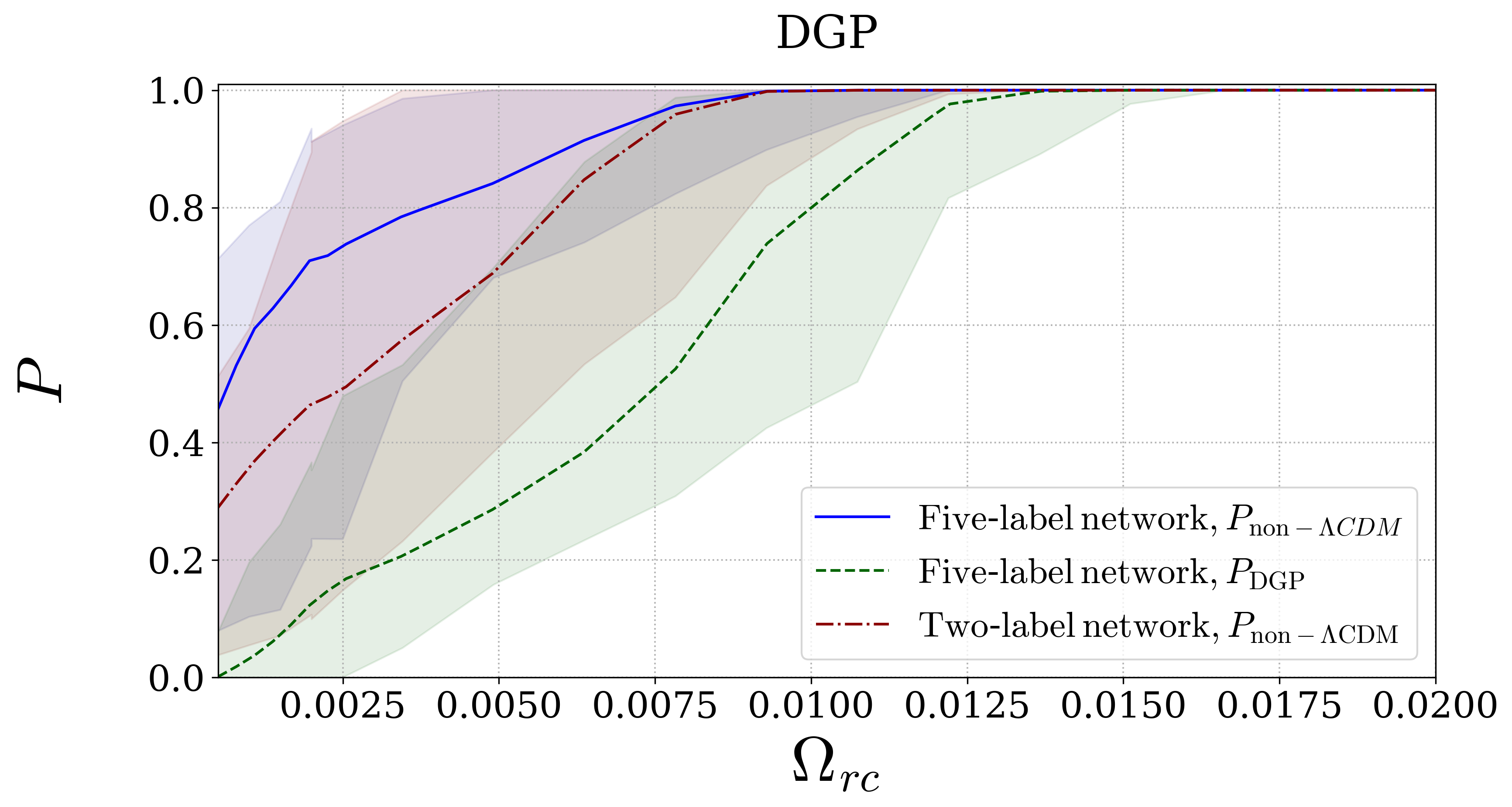}
     }
    \caption{Noise-averaged non-$\Lambda$CDM classification probabilities and associated confidence bands for $f(R)$ (Top) and DGP (Bottom) spectra for the five and two-label networks as a function of $f_{R0}$ and $\Omega_{rc}$. 
    The five-label classification probabilities $P_{f(R)}$ and $P_{\mathrm{DGP}}$ are also shown.
    One can see that the average non-$\Lambda$CDM classification probability for the five-label network provides the most robust indicator of the presence of a modification, confidently classifying spectra as non-$\Lambda$CDM for $f_{R0}\approx 9\times10^{-8}$ and $\Omega_{rc} \approx 0.008$ independently of the noise for $f(R)$ and DGP respectively.
    }
    \label{fig:fR_DGP_constraints}
\end{figure*}
\begin{figure*}
    \centering
    \resizebox{0.49\textwidth}{!}{
    \includegraphics{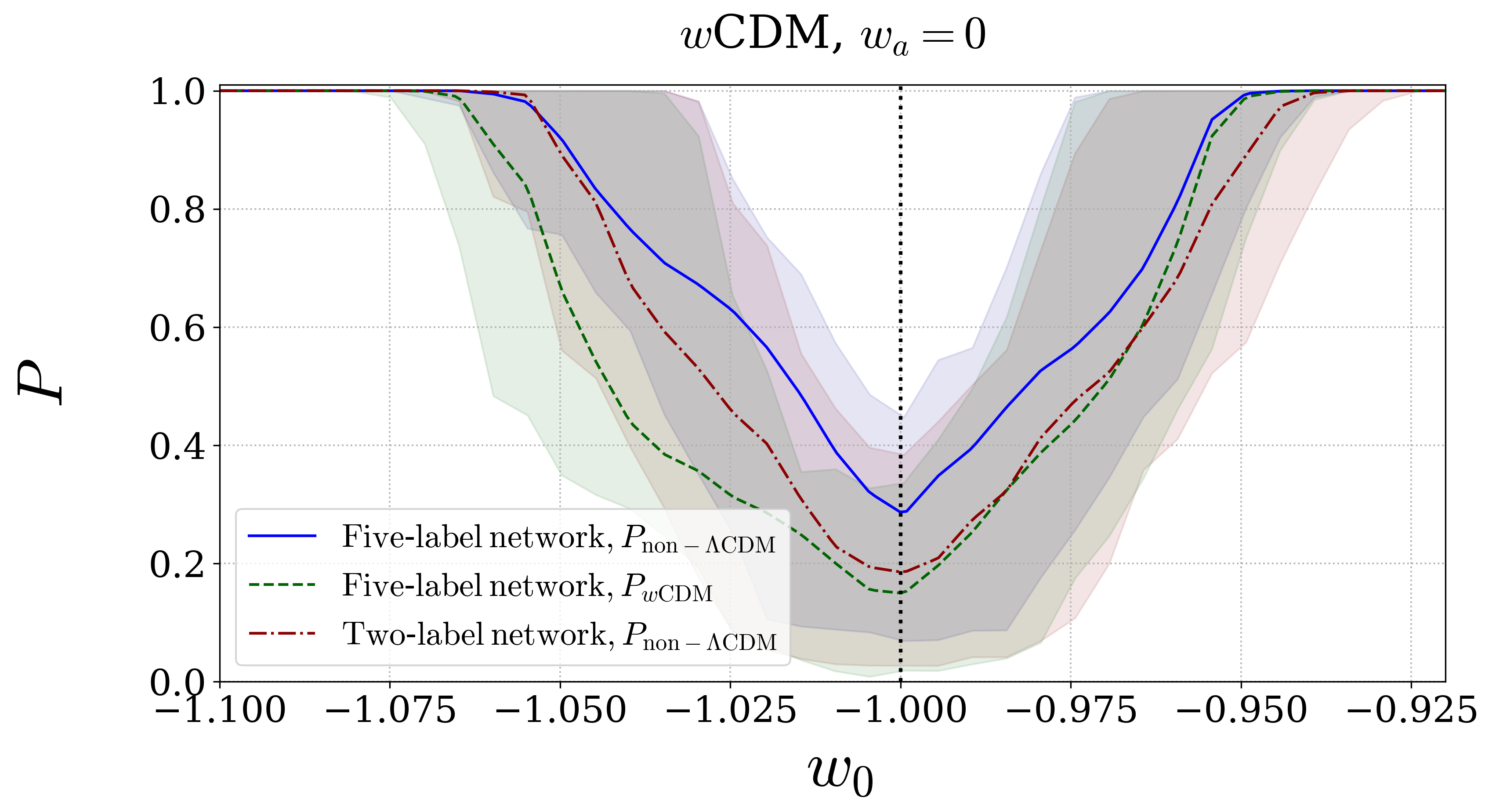}
     }
     \resizebox{0.49\textwidth}{!}{
    \includegraphics{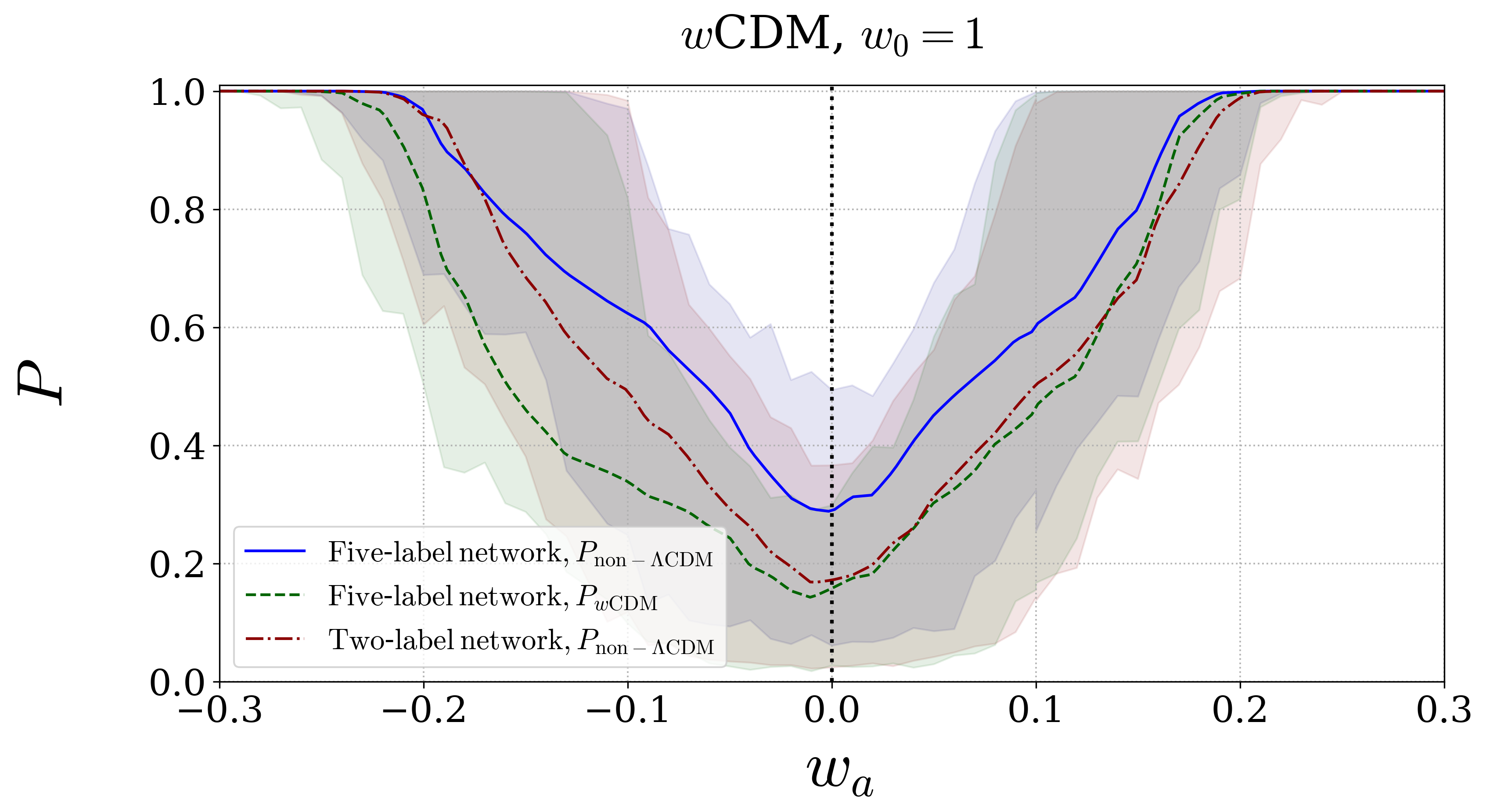}
     }
    \caption{Noise-averaged non-$\Lambda$CDM and $w$CDM classification probabilities for deviations of $w_{0}$ (Top) and $w_{a}$ (Bottom) around their fiducial values. 
    On average, the five-label network is better at detecting deviations from $\Lambda$CDM in each case. 
    However the performance of each network remains highly sensitive to the noise in the ranges $-1.07\lesssim  w_{0} \lesssim -0.94$ and $-0.25\lesssim  w_{a} \lesssim 0.25$.
    }
    \label{fig:wcdm_constraints}
\end{figure*}
In Figs.~\ref{fig:fR_DGP_constraints} and \ref{fig:wcdm_constraints} we present the results for $f(R)$, DGP and $w$CDM. 
One can see that in the case of $f(R)$ gravity the five-label network's non-$\Lambda$CDM classification probability is more capable of recognising small deviations in $f_{R0}$, on average classifying spectra as non-$\Lambda$CDM when $f_{R0}>8\times 10^{-8}$. 
The same network becomes more confident that a spectrum specifically belongs to $f(R)$ for $f_{R0}>1\times 10^{-7}$.
Conversely, the two-label network's ability to confidently classify spectra as non-$\Lambda$CDM remains highly sensitive to the noise up to $f_{R0}> 1.4 \times 10^{-7}$.

In the case of DGP, while the five-label network's non-$\Lambda$CDM classification probability again provides the most reliable predictions, the two-label network's ability to classify spectra as non-$\Lambda$CDM outperforms the five-label network's ability to classify the spectra as DGP.
For values of $\Omega_{rc}>0.016$ both networks definitively determine all spectra are not $\Lambda$CDM independently of the noise. 

Turning our attention to  each network's ability to detect evolving dark energy, we show in Fig.~\ref{fig:wcdm_constraints} the noise-averaged classification probabilities for a range of $w$CDM power spectra.
In each case we vary either $w_{0}$ or $w_{a}$ fixing the non-varying parameter to their $\Lambda$CDM fiducial values of $\left(w_{0},w_{a}\right) = \left(-1,0 \right)$.
Again we find that for both $w$CDM parameters, the five-label non-$\Lambda$CDM classification probability is the most reliable indicator of a deviation from $\Lambda$CDM. 
Despite both networks on average classifying $w$CDM as non-$\Lambda$CDM for deviations of $\Delta w_{0}\sim 0.05$ and $\Delta w_{a} \sim 0.2$, the five-label network is less sensitive to the noise. 
We leave a detailed analysis of how the degeneracies between $w_{0}$ and $w_{a}$ affect the noise-averaged classification probability to future work.  
\begin{figure*}
    \centering
    \resizebox{0.5\textwidth}{!}{
    \includegraphics{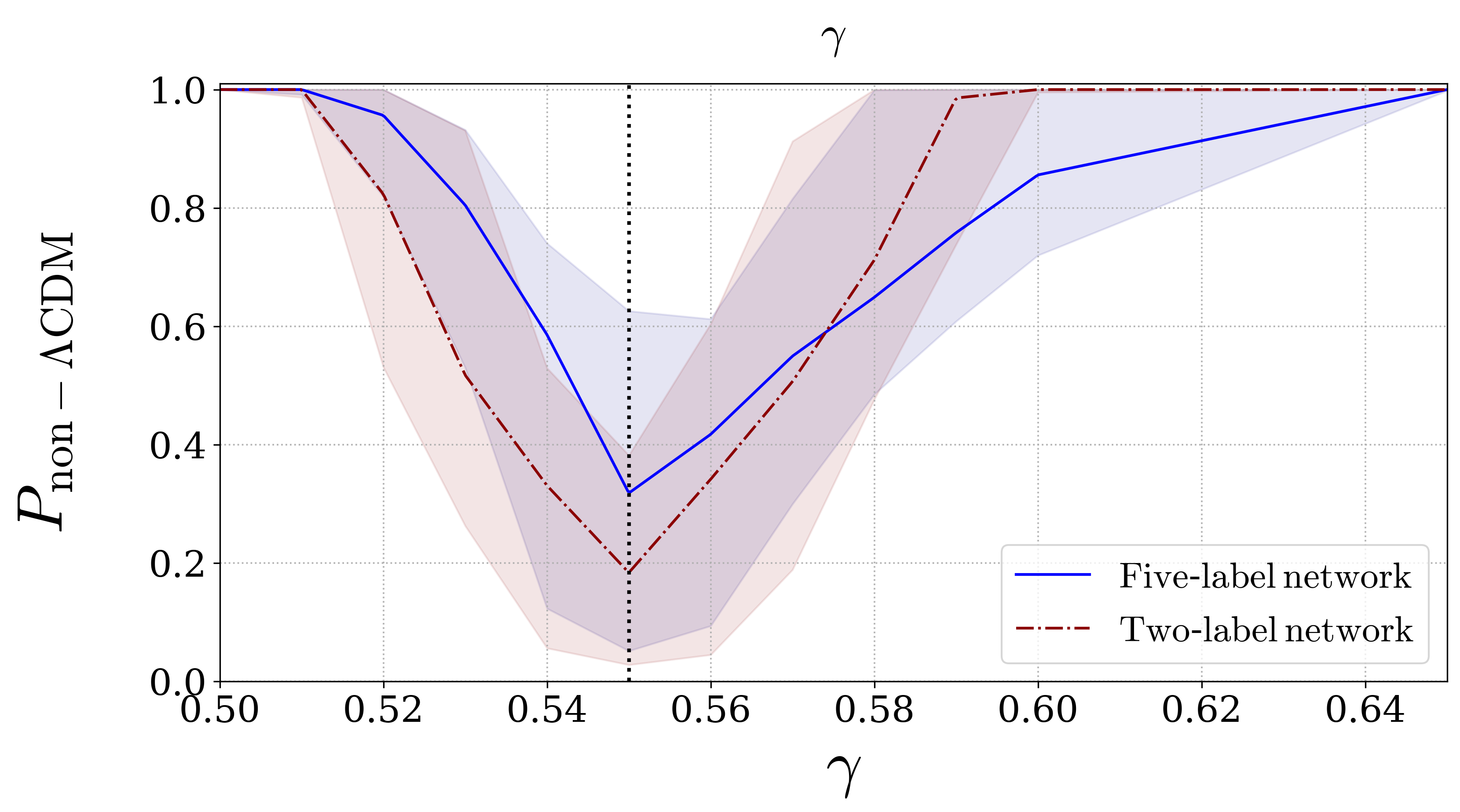}
     }
    \caption{Noise-averaged five-label and two-label classification probability for spectra generated using the growth-index $\gamma$.
    For values of $\gamma>0.6$ the two-label network classifies the spectrum as non-$\Lambda$CDM with a much higher confidence than the five-label network.
    However, we attribute this to the fact that the five-label network is better able to determine that this spectrum is an out-of-distribution example.
    }
    \label{fig:gamma_constraints}
\end{figure*}
\begin{table*}
\centering
\begin{tabular}{ | c | | c | c | c | c | c |}
 \cline{2-6}
 \multicolumn{1}{c | }{} & \multicolumn{5}{c |}{Extensions} \\ \hline 
 Parameter &  $w_0$ & $w_a$ & ${f}_{\rm R0}$ &  $\Omega_{\rm rc} $ & $\gamma$\\  \hline 
 Five-label & $\left(-1.05,-0.95 \right)$ & $\left(  -0.20,0.17\right)$ & $8 \times 10^{-8}$ & 0.007 & $\left(0.52, 0.63\right)$\\ \hline 
 Two-label   &  $\left(-1.05, -0.95 \right)$ &  $\left(-0.20,  0.19\right)$  & $10^{-7}$ & 0.008 & $\left(0.51, 0.59\right)$\\ \hline
\end{tabular}
\caption{Values of the minimum deviation in each of the model parameters such that both the five-label and two-label networks classify a spectrum as non-$\Lambda$CDM at $95\%$ confidence when averaged over Gaussian noise realisations.
In the case of $w$CDM, these bounds are obtained by fixing either $w_{0}$ or $w_{a}$ to their fiducial value and allowing the other to vary. 
We refer the reader to Figs.~\ref{fig:fR_DGP_constraints},~\ref{fig:wcdm_constraints} and \ref{fig:gamma_constraints} for estimates of the variance in the classification probability for each of these values.  
}
\label{tab:param_constraints}
\end{table*}

Having completed these tests for models belonging to the training set, we repeat the procedure for spectra generated with varying values of the growth-index $\gamma$ as outlined in Sec.~\ref{sec:out_of_dist}.

In Fig.~\ref{fig:gamma_constraints} we show the noise-averaged classification probability for deviations around the $\Lambda$CDM fiducial value of $\gamma=0.55$.
Due to the absence of a specific label for $\gamma$, in this case the lower and upper bounds for the confidence bands are constructed by selecting the noise realisations that maximise and minimize $\mu_{\Lambda \mathrm{CDM}}$ respectively.
Although it appears that the two-label network can pick up smaller deviations than the five-label network for $\gamma>0.55$, in this case it is because the five-label network recognises the spectrum does not belong to the models in the training set and thus assigns a lower classification probability. 
For values of $\gamma < 0.55$ the spectra are more degenerate with DGP which may help each network recognise the spectrum as non-$\Lambda$CDM. 
Even in this instance however, the probability is split between DGP and other non-$\Lambda$CDM labels as in Fig.~\ref{fig:gamma_5label}.
Note that the bounds on the growth index $\gamma$ were obtained from a network which was not trained on spectra generated from $\gamma$.
One would therefore expect these bounds to improve if such spectra were included in the training set.

\subsection{Impact on future experiments}
\label{sec:experiments}
In Table~\ref{tab:param_constraints} we summarise the values of the minimum magnitude of each model parameter for every $\Lambda$CDM extension such that the noise-averaged non-$\Lambda$CDM classification probability is approximately $95\%$ for both the five-label and the two-label network. 
The uncertainty estimated in our approach to the BNN output is of similar order of magnitude of stage IV astronomical survey forecasts, despite the two methods not being directly comparable, nor do we use the observational probes of upcoming surveys, e.g.~weak lensing and galaxy clustering.
We note the $1\sigma$ cosmic shear forecasts of Ref.~\cite{Bose:2020wch} were $f_{\rm R0} \leq 10^{-7.25} $ and $\Omega_{\rm rc} \leq 0.08 $ which assumes an LSST-like survey and a multipole scale cut of $\ell_{\rm max} = 1500$.
The official Euclid Fisher forecast of Ref.~\cite{Blanchard:2019oqi} gives $w_0 = -1 \pm 0.097 (0.077) $ and $w_a =0 \pm 0.32 (0.24) $, which combines both galaxy clustering and weak lensing probes and pessimistic (optimistic) scale cuts. 
The $2\sigma$ constraints estimated for Euclid on $\gamma$ are $\gamma = 0.55 \pm 0.036 (0.026)$ for the pessimistic (optimistic) analyses of Ref.~\cite{Blanchard:2019oqi} for WL+${\rm GC}_{\rm s}$.
This means that the method outlined here is able to pinpoint deviations from $\Lambda$CDM down to a level relevant for Euclid.

\section{Training specialist networks}
\label{sec:specialist}

\begin{figure*}
    \centering
    \resizebox{0.5\textwidth}{!}{
    \includegraphics{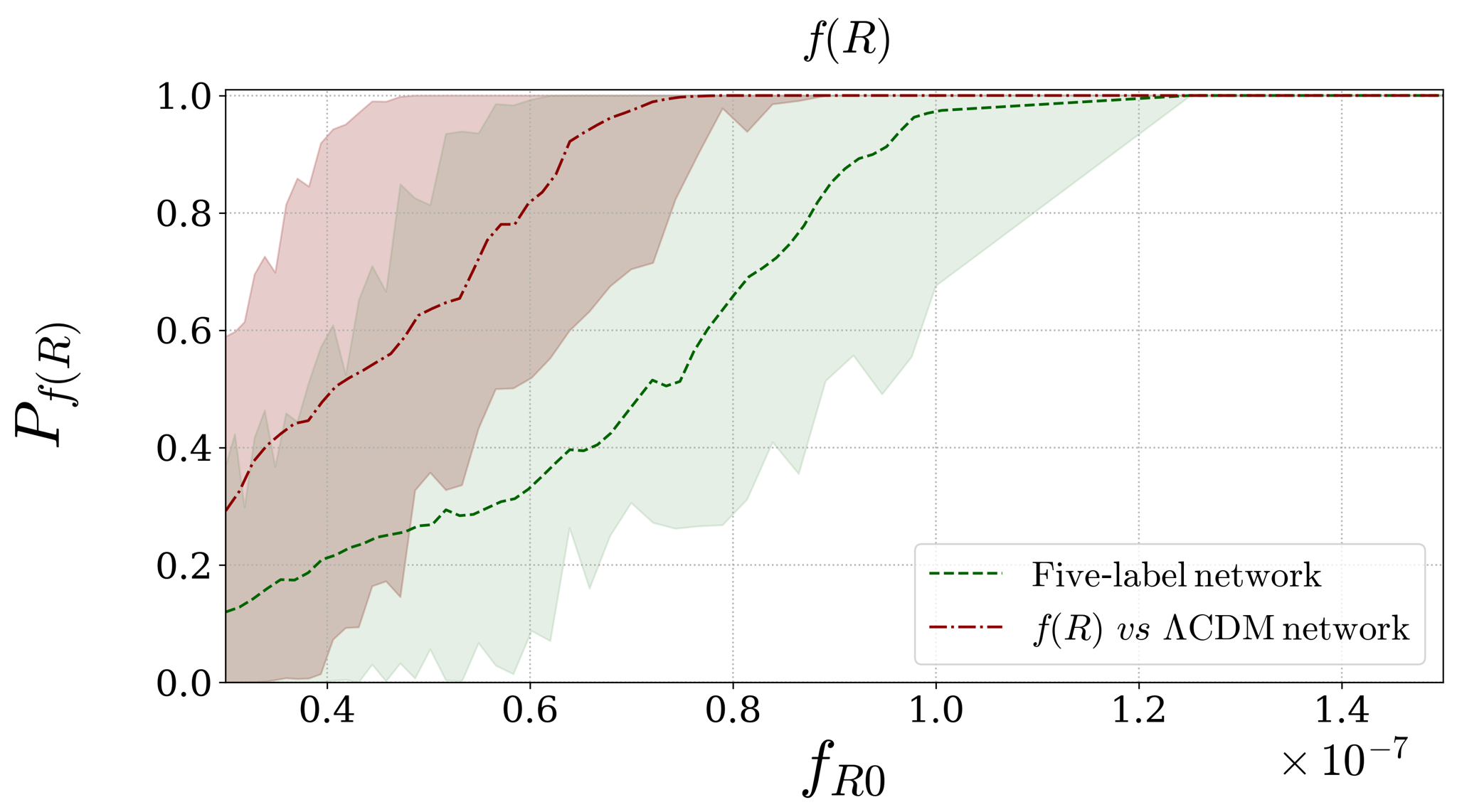}
     }
    \caption{We compare how the ability of both the five-label network and a specialist $f(R)$ network to correctly classify $f(R)$ spectra varies with the modification strength $f_{R0}$ when averaged over noise realisations.
    It is clear that the specialist network outperforms the five-label network, with $f(R)$ spectra with $f_{R0} > 8 \times 10^{-8}$ being correctly classified largely independently of the noise realisation.
    }
    \label{fig:noise_avg_comparison}
\end{figure*}

Given the promising performance of the five-label network in detecting deviations in $f(R)$ models down to $f_{R0}\approx \mathcal{O}(10^{-8})$, in this section we discuss the potential gains that could be achieved by training additional networks on subsets of the original five classes in the training set.
Heuristically this follows the philosophy of an MCMC analysis in that in order to constrain a specific model it is beneficial to choose the most appropriate set of model parameters in the MCMC.
In the case of BNNs, if one is only interested in constraining a single model beyond $\Lambda$CDM then in order to maximise the performance of the BNN it is beneficial to only train the network on this model alongside $\Lambda$CDM. 
Such a network would be ``specialised'' to pick up any deviation from the particular source of new physics one is interested in, at the expense of losing information on potential degeneracies between different models when trained on multiple theories.
In this subsection we discuss one such specialist network trained on 18,475 $f(R)$ and 18,475 $\Lambda$CDM power spectra, each of which is passed to the network during the training and validation process with ten different realisations of the noise for a total training and validation set of 369,500 power spectra. 
We use the architecture displayed in Fig.~\ref{fig:model_architecture} where the final layer is now a binary classifier for the two new labels $f(R)$ and $\Lambda$CDM, finding that the training and validation accuracies reach approximately 99.5\%, exceeding that of the five-label network.
We now study how capable this specialist network is in constraining $f_{R0}$ in comparison with the five-label network. 
In Fig.~\ref{fig:noise_avg_comparison} we display a plot of how the noise-averaged $f(R)$ classification probability varies for power spectra generated with values of $f_{R0} \in \left[3\times10^{-8},1\times10^{-7}\right]$ with associated confidence bands.
We find that the performance of the specialist $f(R)$ network exceeds that of the five-label network's $f(R)$ classification capability, retaining a noise-averaged detection confidence of $1\sigma$ for $f_{R0} \approx 5.5\times 10^{-8}$ where the equivalent noise-averaged detection probability for the five-label network is $<0.3$.  %
Furthermore, for values of $f_{R0}> 8\times 10^{-8}$ the classification probability for the specialist BNN asymptotes to one with only a few noise realisations decreasing this probability to $85\%$. 
Spectra with values of $f_{R0}> 9\times 10^{-8}$ are classified at high confidence regardless of the noise realisation.   
By contrast, the five-label network correctly classifies spectra independently of the noise when $f_{R0}> 1.2 \times 10^{-7} $.
Given the limited performance of the generic two-label $\Lambda$CDM vs non-$\Lambda$CDM network and the enhanced performance of the specialised $\Lambda$CDM vs $f(R)$ network we conclude that training a two-label network is principally beneficial when trained between well-defined physical models. 
Tighter constraints on model parameters can also be attained with such a specialised network.

\section{Outlook}\label{sec:outlook}
The potential of BNNs to provide insights into whether upcoming cosmological datasets contain signatures of physics beyond $\Lambda$CDM motivates further exploration.  
For example, BNNs could be used to identify high priority areas in the theory space by selecting the most likely known theory or motivating the need for further model development, before performing parameter estimation and Bayesian model selection with standard techniques that require specific benchmarks from which deviations can be detected.
Indeed, our analysis of how the five-label BNN classified an example generated from the growth-index $\gamma$ in Fig.~\ref{fig:gamma_5label} demonstrates the advantages of using data from specific models over more generic parameterisations. %

With enough training examples from enough $\Lambda$CDM and non-$\Lambda$CDM power spectra generated from a larger set of model classes considered in this work, one may envision a sequence of pre-trained specialist networks such that the first is trained on as many deviations as possible from $\Lambda$CDM, with the latter networks trained on smaller subsets of the total number of classes. 
When an unknown spectrum is passed to the first network it would determine which of the subsequently more specialised networks to pass the spectrum onto.
As the specialist networks are better at recognising the specific imprints of the models they are trained on, they would further classify the spectra until it falls into either a single class or is not confidently classified.
If a single model is indeed preferred following such an analysis, one could then proceed to constrain the model parameters, for example with a traditional MCMC analysis.   
An additional advantage is that, once the network has been trained, one has a tool to rapidly indicate the presence of new physics in contrast to the many hours it would take to obtain constraints with MCMC which must be run on a theory-by-theory basis.  

Importantly however, an MCMC possesses a well defined notion of confidence such that, for a given dataset and a given parameterisation it will converge to a unique set of confidence intervals for each parameter if allowed to run for a sufficient amount of time. Moreover, it allows the computation of quantities such as the Bayesian evidence, that have a well-defined interpretation in terms of model selection and a solid statistical ground, albeit being non-trivial to compute in practice \cite{Heavens:2017afc}.
As we have seen, defining a classification confidence from the output of a BNN can prove challenging as it is by no means trivial to account for the uncertainty arising in the training process, the noise in the data or the chosen network architecture. In particular, the notion of confidence introduced in this paper ensures that the resulting classification is not ``over-confident'' by encoding the uncertainty due to the noise. 
However, we stress that this quantity is not directly comparable to the Bayesian evidence or other goodness-of-fit tests. If and how these notions are comparable remains an open question (see App.\ref{app:MCMC}).
In conclusion, performing a fair comparison between the two methods is not a straightforward endeavour and a more thorough analysis of their relative strengths and weaknesses in performing cosmological analyses will be a subject of future work. 
For now, we see the use of our BNN as a supplementary and precursory tool to MCMC analyses that accelerates the search for new fundamental physics.

While we have restricted ourselves to the matter power spectrum in this work, it is not a directly observable quantity.
An additional study would therefore be to train BNNs on mock galaxy clustering and weak lensing data for a range of different theories to determine how capable BNNs are in detecting deviations from $\Lambda$CDM directly from observational data.
An interesting study in this context was performed in Ref.~\cite{Peel:2018aei}, using a convolutional neural network trained on simulated convergence maps.
Moreover, while we have restricted to a selected number of popular extensions to $\Lambda$CDM, this process is generically applicable to any non-standard theory for which it is possible to rapidly generate accurate power spectra.
Of particular interest would be to study the capability of BNNs to pick up signatures in the power spectrum from the presence of massive neutrinos, baryonic effects as well as modifications arising from Horndeski scalar-tensor theory. 
In the case of massive neutrinos, training a specialist massive neutrino network in a similar manner to that performed for $f(R)$ gravity in Fig.~\ref{fig:noise_avg_comparison} could yield an estimate on how capable BNN's could be in indicating the presence of a non-vanishing neutrino mass. This could be achieved with the latest version of {\tt ReACT} \cite{Bose:2021mkz}.
With the ever-growing ability to model cosmological observables for a multitude of extensions to $\Lambda$CDM, in Sec.~\ref{sec:specialist} we discussed the possibility of training a hierarchy of increasingly more specialist networks to obtain a more confident classification for a spectrum belonging to an unknown class.
Note that one could also fine-tune an $N$-label BNN to distinguish between sub-classes by incorporating an additional layer on top of a previously trained $N$-label network and re-train on a smaller dataset containing the new labels. 
Although training a two-label network from scratch takes a few hours on a GPU, we have included the option in $\tt{BaCoN}$ to fine-tune on new data in the event one wishes to adapt the BNN to classify new theories with a limited training set size.

With the capability of BNNs to extract the particular features in the data that were important in the resulting classification, it may also be possible to provide information on the length scales or redshift bins which should be probed to detect signatures of a particular theory.   
One can then train a BNN on spectra from specific redshift or scale bins.
This may be especially useful for studying models such as $w$CDM whose signatures can be noise-dominated at large scales and low redshift. 
It is also important to note that in this paper we have restricted to a $k$ range of $\left(0.01 - 2.5\right)h \mathrm{Mpc}^{-1}$.
We expect that the capability of BNNs to accurately classify models such as $f(R)$ will only increase with improvements in the ability to rapidly and accurately model power spectra at higher values of $k$. 

Further avenues of exploration include studying the potential benefits of different choices of network architecture and hyperparameters. 
It is also of interest to examine different methods of constructing probability distributions to account for various sources of uncertainty in the output of the BNN. 
In this paper we have focused on the application of BNNs to a classification problem. 
However, the question remains of how capable Neural Networks are in obtaining cosmological parameter constraints from unknown spectra in comparison with more traditional approaches such as MCMC. 
Finally, although we have trained the network on data which lies within the bounds of Euclid errors, it is important to determine how effective BNNs could be in detecting new physics from other surveys such as LSST \cite{Ivezic:2008fe} or DES \cite{Troxel:2017xyo} as well as to investigate how the results vary with different choices of systematic error.

\section{Conclusions}
\label{sec:conclusions}

Over the coming years many new cosmological surveys will provide vast datasets which will determine whether $\Lambda$CDM remains concordance cosmology.
In this paper we have studied the ability of Bayesian Neural Networks to determine if a matter power spectrum is representative of $\Lambda$CDM or not.
By constructing a mapping from the output of a BNN to a well-defined probability distribution we were able to define a classification confidence for an individual spectrum that considers the uncertainty from the noise in the data, variations in the training procedure, the modelling uncertainty of the BNN and choice of hyperparameters. 
We found that a five-label network trained to classify between $\Lambda$CDM, $f(R)$ gravity, DGP gravity, $w$CDM and a ``random'' class provided more reliable predictions than a two-label network trained to distinguish simply between $\Lambda$CDM and non-$\Lambda$CDM.
While generally being less sensitive to variations in the noise distribution, it can also determine whether a power spectrum does not belong to any class included in the training set. 
Since the selection of the correct model is crucial when performing conventional statistical analyses such as with MCMCs, this ability could prove beneficial in indicating prospective models to consider.
However, the network used in this work is currently limited to classification tasks while the notion of model selection on firm statistical grounds in the context of BNNs remains an open problem.
Nevertheless, we found that when averaged over noise realisations the five-label BNN was able to recognise spectra as not being $\Lambda$CDM down to values of $f_{R0} \lesssim 10^{-7}$, $\Omega_{rc} \lesssim 10^{-2} $, $-1.05 \lesssim w_0 \lesssim 0.95 $, $-0.2 \lesssim w_a \lesssim 0.2 $, $0.52 \lesssim \gamma \lesssim 0.59 $, all of which are comparable with current forecasts, as discussed in Sec. \ref{sec:experiments}. 
Specialist networks trained on specific subsets of the classes in the training set have the potential to improve such bounds even further. 

We conclude that BNNs may provide a powerful new means to search for hints of new physics in cosmological datasets.
In particular, we anticipate they will serve as a powerful ``filter'', allowing us to narrow down the theory space before moving on to constrain model parameters with MCMCs while perhaps even signalling the presence of new physics that does not belong to any known model.

\section*{Acknowledgements}
We thank Tom Charnock for useful discussions. We thank Gregory Horndeski for suggesting and allowing the use of \emph{Blustery Mountain Road on an Autumn Day} in this work. 
J.K., B.B., and L.L.~acknowledge the support by a Swiss National Science Foundation
(SNSF) Professorship grant (No.~170547). 
The work of M.M.~is supported by the SwissMap National Center for Competence in Research.
BNN development was conducted on Google Colab.
Please contact the authors for access to research
materials.

\section*{Data and code availability}
Alongside this paper we publish the publicly available code Bayesian Cosmological Network ($\tt{BaCoN}$) which can be accessed at the github repository \url{https://github.com/Mik3M4n/BaCoN} with the training and test data available at \url{https://doi.org/10.5281/zenodo.4309918}.

\bibliography{mybib}

\appendix

\section{Neural networks for supervised learning}\label{app:NNintro}

In this appendix we introduce some basic concepts on neural network classifiers.
Consider a dataset of the form $\mathcal{D}=\{(X, y)_k\}, \, k\in [1, ... , M]$ where each element consists of a pair of \emph{features} $X$ and an associated label $y$ and $M$ denotes the size of the dataset. 
Let us further denote $N$ to be the number of possible labels.
In a supervised classification task the aim is to use the labelled examples in $\mathcal{D}$ in such a way to be capable of predicting $y^{\star}$ for a previously unseen $X^{\star}\notin \mathcal{D}$.
Note that for our purposes, each $X_{k}$ consists of a collection of matter power spectra in different redshift bins with $y_{k}$ labelling the associated underlying physical model (see Sec.~\ref{sec:datagen}).

Neural networks provide a powerful means to model nonlinear features in labelled data by combining a hierarchy of nonlinear functions in a succession of \emph{layers}, each with optimised parameters, which map given features to a predicted label.
Different choices can be made for the number of layers, the type and size of each layer, all of which constitute the network's \emph{architecture}.
To begin, the labelled dataset is split into a \emph{training set}, \emph{validation set} and a \emph{test set}. 
By passing the training data through the network, usually in a series of batches, the network parameters are tuned using an optimization algorithm to minimise a \emph{loss function} which quantifies how close the output of the network matches the associated label of the input data. 
An \emph{epoch} occurs when every batch of data in the training set has been passed through the network.

Central to the optimization procedure is \emph{gradient descent} which updates the parameters in the direction where the derivative of the loss, computed using the \emph{backpropagation} algorithm, is maximally negative.
Many modifications to gradient decent have been developed in order to aid the optimization efficiency and the choice of optimisation algorithm and its associated parameters, known as \emph{hyperparameters}, is an important factor in determining the performance of the network. 
In this paper we use the \emph{adam} optimisation algorithm \cite{kingma2017adam}.
The most relevant hyperparameter is the \emph{learning rate} which sets the amplitude of the step made in the direction of the gradient.

Following each epoch a performance metric is computed to evaluate how effectively the network maps features to labels in both the training set and the validation set, the latter giving a measure on how well the network generalises to previously unseen data. 
Note that this metric need not be the same function as the loss. 
In particular, the loss function must be differentiable with respect to the weights while the metric does not. 
Different choices for the loss function and the metric depend on the problem at hand and are a key consideration in the network design. 
Typically in classification problems the performance metric is the accuracy which is simply the fraction of correctly classified examples. 
This process is then repeated until the loss stabilises to a minimum. 
As the performance metric on the validation set can remain biased, the final stage is to evaluate the network performance on the test set which has not been used in the training process. 
If the performance metric on the test set is comparable to that on the training and validation sets then one can be confident the network is robust. 
%

%
\section{Classification in BNNs}\label{app:BNNtrain}
Evaluating the performance of a neural network cannot be limited to evaluation of a performance metric on the test set, especially if the network is to be used in a scientific context. 
In this case it is imperative to assess its reliability on any individual prediction and to define a probability that quantifies how much the prediction can be trusted. 
%

Using traditional DNNs to compute both aleatoric and epistemic uncertainties (defined in Sec.~\ref{sec:BNNs}) would be both computationally expensive and time-consuming. 
This appendix details why this is so before discussing how BNNs are better suited to model classification uncertainties.  
We define the labels to be \emph{one-hot encoded}, such that they are vectors of length N 
with a one at the position of the true label and with zeroes otherwise. 
For example, the vector $y=(1, 0, ..)$ is a label for an example belonging to the first class. 
Classification occurs when the final layer of the network outputs an $N$-dimensional vector with components that sum to one and can therefore be interpreted as probabilities. 
Denoting $f(X |  w, a)$ the vector-valued output of the final layer given the weights $w$ and an architecture $a$, a probability that $X$ belongs to the $i$th class can be obtained by passing it to the \emph{softmax} function
\begin{equation}\label{softmax}
    p(y_{i}=1 | X,  w, a) = \frac{e^{-f_i(X |  w, a)}}{\sum_{i=1}^{N} e^{-f_i(X |  w, a)} } \, .
\end{equation}
We can then choose a multinomial probability distribution as a likelihood such that
\begin{align}
    &\mathcal{L}(\mathcal{D} | w, a) = \prod_{k=1}^{M} \prod_{i=1}^{N}  \left[p(y_{k, i}=1 | X_k, w,  a) \right]^{y_{k, i}} \, , \\ \nonumber
    &\quad \sum_{i=1}^{N} p(y_{k, i}=1 | X_k, w,a) =1 \, ,
\end{align}
with the loss function being the negative log-likelihood. 
From now on we shall drop the explicit dependence on the architecture $a$ 
but it should be kept in mind that all the results are conditioned on the choice of $a$.

The training procedure yields a maximum likelihood estimate set of weights $\hat{w}$.
When predicting the label for a new example with features $X^{\star}$ the network outputs the probability
\begin{equation}\label{prediction}
     p(y_{i}^{\star}=1 | X^{\star}, \hat{w}, \mathcal{D}) \quad \forall i=1, ..., N \, ,
\end{equation}
where the conditioning on $\mathcal{D}$ and $\hat{w}$ indicates that the training has been performed with this dataset resulting in a particular maximum likelihood estimate for the weights. 
Note however that $\hat{w}$ is not a unique value dependent on $\mathcal{D}$ and the optimisation process, due to the inherent stochasticity of the training process.  
A prediction for the label is obtained by assigning the label to the maximum output probability
$y_{\text{pred}}^{\star} = {\arg\max}_i \, p(y_{i}^{\star}=1 | X^{\star}, \hat{w},\mathcal{D}, a)$ if this exceeds a chosen threshold probability $p_{th}$.
One must be careful not to interpret Eq.~\eqref{prediction} as the confidence in the prediction due to the explicit dependence on $\hat{w}$, $\mathcal{D}$, variations in the training procedure, choice of optimisation algorithm or initialisation of the weights, and the presence of aleatoric uncertainty. 
Estimating the uncertainty would require the expensive procedure of averaging the results from an ensemble of independently trained DNNs.

Fortunately, BNNs can quantify the uncertainty more efficiently by replacing each weight in the network by a parameterised distribution \cite{MacKay, Neal}. 
The training objective is then to infer the posterior distribution of the weights conditioned on the training data
\begin{equation}
    p(w | \mathcal{D}) = \frac{\mathcal{L}(\mathcal{D} | w) p(w) }{p(\mathcal{D})} \, .
\end{equation}
In practice, $p(w | \mathcal{D})$ is intractable and so approximations or sampling techniques are employed.
One such approximation approach is variational inference where the posterior is approximated by a variational distribution $q_{\theta}(w)$ which describes a family of distributions parameterised by the parameter $\theta$ \cite{Jordan1999, 10.5555/2986459.2986721, Gal2016Uncertainty, Blei_2017}.
In training the BNN, the objective is to ensure the resulting variational distribution $q_{\theta}(w)$ matches the posterior weight distribution $p(w | \mathcal{D})$ as accurately as possible. 
To achieve this it is necessary to have a measure on the difference between two distributions which could serve as a loss function.
One such measure capable of quantifying how much the two distributions $q_{\theta}(w)$ and $p(w|\mathcal{D})$ differ is the K{\"u}llback-Leibler (KL) divergence given by \cite{kullback1951}
\begin{equation}
    \textnormal{KL}\left[q_{\theta}(w)||p(w|\mathcal{D})\right] \equiv \int dw \,\, q_{\theta}(w) \log\frac{q_{\theta}(w)}{p(w|\mathcal{D})}  \, .
\end{equation}
Using Bayes theorem to re-express the posterior $p(w | \mathcal{D})$ in terms of the likelihood $\mathcal{L}(\mathcal{D}|w)$ and the prior distribution over the weights $p(w)$ this can be re-expressed as \cite{Jaakkola2000, article_Neal_Hinton, Blundell2015}
\begin{align}
    \textnormal{KL}\left[q_{\theta}(w)||p(w|\mathcal{D}))\right]= \, &\textnormal{KL}\left[q_{\theta}(w)||p(w)\right] \nonumber \\
    -& \mathbb{E}_{q_{\theta}(w)}\large[ \log \mathcal{L}(\mathcal{D}|w) \large] + const \, ,
\label{Var_free_energy}    
\end{align}
where the constant term arises from the Bayesian evidence which does not affect the optimisation process.

The KL-divergence between the variational distribution and the prior can be interpreted as a regularisation term that ensures the variational distribution does not become too complex, potentially leading to overfitting, while the second term is the usual negative log-likelihood.
By sampling the weights from the variational distribution $w \sim q_{\theta}(w|\mathcal{D})$ one can obtain a Monte Carlo (MC) estimate for the loss in Eq.~\eqref{Var_free_energy}.
However, given that the weights $w$ are now random variables it is not possible to take derivatives directly to perform gradient descent. 

To circumvent this issue ~\cite{Kingma2013},~\cite{ Kingma2015} detail a reparameterisation trick which, rather than sampling directly from the variational distribution, samples a new random variable $\epsilon$ from a standard Gaussian such that $\epsilon \sim p(\epsilon)$ where $p(\epsilon) = \mathcal{N}(0, 1)$.  
This in turn is related to the weights via a deterministic function such that $w = g(\epsilon, \theta)$.
Now that the weights are expressed as a deterministic function, itself now a function of the random variable $\epsilon$, it is possible to perform backpropagation. 
The drawback of this approach is that the resulting sampled weights are the same for each batch, correlating the resulting gradients and slowing the convergence of the optimisation algorithm.  
In order to decorrelate the gradients across the batch Ref.~\cite{wen2018flipout} proposed the \emph{Flipout} method.
Assuming the variational distribution can be expressed as a mean plus a perturbation, by randomly multiplying each perturbation by either $\left\{ 1, -1\right\}$ one can ensure that the weights across a batch are at least partially decorrelated. 
This method has proven to be effective in recent applications of BNNs \cite{2020PhRvD.102j3509H, Lin:2020aps}, with the additional advantage of being available as pre-built implementations in popular Deep Learning libraries like TensorFlow \footnote{\protect\url{https://www.tensorflow.org/probability/overview}} for both dense and convolutional layers \cite{dillon2017tensorflow}. 
In this paper we make use of TensorFlow \cite{tensorflow2015-whitepaper} and TensorFlow Probability \cite{dillon2017tensorflow} throughout.
Following training, the posterior weight distribution can be used to obtain predictions by marginalising over the weights, generalising Eq.~\eqref{prediction} to
\begin{equation}\label{probabilityBayesExact}
    p(y_{i}^{\star}=1 | X^{\star}, \mathcal{D}) = \int p(y_{i}^{\star}=1 | X^{\star}, w, \mathcal{D}) p(w | \mathcal{D}) dw \, .
\end{equation}
In practise, this equation is evaluated by Monte Carlo sampling from the distribution $q_{\theta}(w)$, yielding Eq.~\ref{probabilityBayes} in the main text.

%
\section{Construction of a probability distribution from the output of a BNN}\label{app:prob}
\subsection{$N$-label distribution}
\label{app:NDgauss}
In this Appendix we shall detail the construction of the probability distribution introduced in Eq.~\eqref{probDist}.
We aim to define a probability for a random variable $x$, with mean $\mu$ and covariance $\Sigma_{q_{\theta}}$. 
The random variable $x$ represents the softmax output of the network.
It is therefore subject to the following conditions that each component of $x$ lies between 0 and 1 and that the components sum to 1, namely
\begin{equation}\label{constraint}
\sum_{i=1}^N x_{i} = 1 \, , \quad \sum_{i=1}^N \mu_{i }=1 \, .
\end{equation}
If the components of $x$ were independent we could define its distribution as a multivariate Gaussian
\begin{equation}\label{multivariate}
\mathcal{N}(x; \mu, \Sigma_{q_{\theta}}) \, ,
\end{equation}
truncated between 0 and 1. 
However, due to the constraint in Eq.~\eqref{constraint} the matrix $\Sigma_{q_{\theta}}$ as defined in Eq.~\eqref{fullCov} is degenerate so that $ \det\Sigma_{q_{\theta}}=0$ implying the multivariate Gaussian distribution is not defined. 
In particular, the columns of the matrix satisfy
\footnote{This and other relations used in this section are a consequence of the fact that Eq.~\ref{fullCov} is the covariance matrix of a multinomial distribution with parameters $\mu_i$.}
\begin{equation}\label{degeneracy}
[\Sigma_{q_{\theta}}]_{Ni} = - \sum_{k=1}^{N-1}[\Sigma_{q_{\theta}}]_{ik} \, ,
\end{equation}
and $\Sigma_{q_{\theta}}$ has a null eigenvalue. 
To circumvent this problem we can introduce a small perturbation $\epsilon$ such that Eq.~\eqref{constraint} becomes
\begin{equation}\label{constraint1}
\sum_{i=1}^N x_{i} = 1-\epsilon \, , \quad \sum_{i=1}^N \mu_{i }=1-\epsilon \, .
\end{equation}
Defining $\tilde{\Sigma}_{q_{\theta}}$ to be the covariance obtained from the definition in Eq.~\eqref{fullCov} with the perturbed constraint \eqref{constraint1}, the degeneracy condition \eqref{degeneracy} becomes
\begin{equation}\label{degeneracy1}
[\tilde{\Sigma}_{q_{\theta}}]_{Ni} = - \sum_{k=1}^{N-1}[\tilde{\Sigma}_{q_{\theta}}]_{ik}+ \epsilon \mu_i \, .
\end{equation}
The matrix $\tilde{\Sigma}_{q_{\theta}}$ is now invertible and the distribution in Eq.~\eqref{multivariate} is well-defined. 
Now we take the limit $\epsilon \rightarrow 0$. 
Since $\Sigma_{q_{\theta}}$ is symmetric, there exists an orthogonal matrix $B$ such that $\Sigma_{q_{\theta}} = B U B^{-1}$, with $U=\text{diag}(u_1,...,u_{N-1}, 0 )$, where $u_i$ are the eigenvalues of $\Sigma_{q_{\theta}}$, $ u_i \neq 0 \; \forall i=1,.., N-1$. 
The effect of the correction $\epsilon$ is to shift the value of the eigenvalues. 
In particular, the last eigenvalue becomes non-zero resulting in the following diagonal matrix
\begin{equation}
\tilde{U}=\text{diag}(u_1 + \alpha_1 \epsilon ,...,u_{N-1}+ \alpha_{N-1} \epsilon, \alpha_N \epsilon ) \, , 
\end{equation}
where the form of the coefficients $ \alpha_1, ...,  \alpha_{N}$ is not relevant to the present discussion. 
Defining $\tilde{B} $ to be the matrix such that $ \tilde{\Sigma}_{q_{\theta}} = \tilde{B} \tilde{U} \tilde{B}^{-1}$, the variable 
\begin{equation}\label{Zdef}
 \tilde{Z} = \tilde{B}^{-1} (x-\mu) \, , 
\end{equation}
is distributed as
\begin{align}\label{multivariateDiag}
\mathcal{N}( \tilde{Z}; 0, \tilde{U}) & =\Bigg( \prod_{i=1}^{N-1} \mathcal{N}( \tilde{Z}_i; 0, \tilde{U}_{ii})\Bigg) \times \mathcal{N}( \tilde{Z}_N; 0, \sqrt{\alpha_N \epsilon}) \nonumber \\ & \equiv \tilde{\mathcal{F}}(\tilde{Z}; 0, \tilde{U}) \, .
\end{align}
In the limit $\epsilon \rightarrow 0$ we have: 
\begin{eqnarray}\label{multivariateDiag_lim}
\tilde{\mathcal{F}}( \tilde{Z}; 0, \tilde{U})& =  \Bigg( \prod_{i=1}^{N-1} \mathcal{N}( \tilde{Z}_i; 0, \tilde{U}_{ii})\Bigg) \times \mathcal{N}( \tilde{Z}_N; 0, \sqrt{ \alpha_N \epsilon}) \nonumber \\
 &\rightarrow \Bigg( \prod_{i=1}^{N-1} \mathcal{N}( Z_i; 0, u_{i})\Bigg) \times \delta\Big(Z_N\Big) \, . 
\end{eqnarray}
Hence, we can define the distribution of $Z\equiv B^{-1} (x-\mu)$ as 
\begin{equation}\label{PdefZ}
\mathcal{F}(Z; 0, U) = \lim_{\epsilon \to 0} \tilde{\mathcal{F}}( \tilde{ Z}; 0, \tilde{U} ) = \delta \Big( Z_N \Big) \times \prod_{i=1}^{N-1} \mathcal{N} ( Z_i ; 0, u_i ) \, .
\end{equation}
From the definition in Eq.~\eqref{Zdef}, it follows the distribution of $x$ can be defined as 
\begin{align}\label{Pdef}
\mathcal{F}(x; \mu, \Sigma_{q_{\theta}}) &= \lim_{\epsilon\to 0} \tilde{\mathcal{F}}(\tilde{B}^{-1} (x-\mu); 0, \tilde{B}^{-1} \tilde{\Sigma}_{q_{\theta}} \tilde{B}) \nonumber \\
&= \delta\Big(\left[ B^{-1} (x-\mu)\right]_N\Big) \nonumber \\ &\times \prod_{i=1}^{N-1} \mathcal{N}\Big( \left[B^{-1} (x-\mu)\right]_i; 0, \left[B^{-1} \Sigma_{q_{\theta}} B\right]_{ii} \Big)   \, .
\end{align}
It can be shown that the matrix $B$ has elements \cite{withers2014}
\begin{align}
B_{ij} &={M}_i (u_j - {M}_i)^{-1} \gamma_j \; , \nonumber \\  \gamma_j &\equiv\Big[ \sum_{k=1}^N {M}_k^2 ({M}_k-u_j)^{-2} \Big]^{-1/2} \, ,
\end{align}
where we remind the reader that $u_j$ is the $j$-th eigenvalue of $\Sigma_{q_{\theta}}$. 
Therefore, for the $N$-th element of the vector $Z$ which has a zero eigenvalue, we have
\begin{align}
Z_N =& \left[ B^{-1} (x-\mu)\right]_N= \sum_{j=1}^N B_{jN} (x_j -\mu_j)  \\ \nonumber = & \sum_{j=1}^N (\mu_j-x_i) \gamma_N  \\ \nonumber 
=&\frac{1}{\sqrt{N}} \Big( 1-\sum_{j=1}^N x_j \Big) \; .
\end{align}
Substituting the above relation into Eq.~\eqref{Pdef} and enforcing the requirement that each value of $x_i$ must lie between 0 and 1, one finally obtains Eq.~\eqref{probDist}. 
Formally, this is accomplished by multiplying the distribution \eqref{Pdef} by a multi-dimensional indicator function of the interval $[0, 1]$ and  properly renormalising, which yields the truncated gaussian distribution denoted by $\tilde{\mathcal{N}}$ in \ref{probDist}.
In practice, we are interested in sampling from the distribution in order to compute the probability $P_{I}$ in Eq.~\eqref{pGauss}. 
In order to draw the samples we proceed with the following algorithm
\begin{algorithmic}[ht]
\While {$n_{samples}\leq goal $ }
    \State \text{- draw samples $Z_i \, , i=1, ... N-1 $,  from  } $\mathcal{N}(0,u_i)$
    \State \text{- set} $Z_N$=0
    \State \text{- Compute } $x=B Z+\mu $
\If {$0< x_i < 1 \; \forall i$}
    \State \text{accept sample}
\Else
\State \text{reject sample}
\EndIf
\EndWhile
\end{algorithmic}
where $goal$ is the desired number of samples and $n_{samples}$ is the total number of valid samples obtained at each step.

\subsection{Two-label distribution}\label{app:2Dgauss}
It is instructive to consider the special case of $N=2$, where the derivation in App.~\ref{app:NDgauss} results in a simple analytic closed form. 
In particular, we find that
\begin{equation}
\mu = (\mu_1, \mu_2) \; , \quad \mu_2=1-\mu_1 \, ,
\end{equation}
\begin{align}
\tilde{\Sigma}_{q_{\theta}}= &
\begin{pmatrix}
\sigma^2 & -\sigma^2+\mu_1 \epsilon\\
-\sigma^2+\mu_1 \epsilon & \sigma^2+\epsilon (1-2 \mu_1-\epsilon)
\end{pmatrix} \; , \nonumber \\  \sigma^2 \equiv& \mu_1-\mu_1^2 \, ,
\end{align}
\begin{align}
\tilde{B}=\frac{1}{\sqrt{2}} 
\begin{pmatrix}
-1+\frac{(1-2\mu_1)}{2\sigma^2} \epsilon+ \mathcal{O}(\epsilon^2) & 1+\frac{(1-2\mu_1)}{2\sigma^2} \epsilon+ \mathcal{O}(\epsilon^2) \\
1& 1
\end{pmatrix} \, ,
\end{align}
\begin{equation}
\tilde{U}=
\begin{pmatrix}
2\sigma^2+(1-2\mu_1) \epsilon+ \mathcal{O}(\epsilon^2) & 0\\
0 &  \epsilon/2+ \mathcal{O}(\epsilon^2) 
\end{pmatrix} \, ,
\end{equation}
and
\begin{equation}
\tilde{Z}=\frac{1}{\sqrt{2}}
\begin{pmatrix}
-X_1+X_2+\mu_1-\mu_2 +\mathcal{O}(\epsilon)\\
X_1+X_2-\mu_1-\mu_2 +\mathcal{O}(\epsilon)
\end{pmatrix} \, .
\end{equation}
After applying Eq.~\eqref{Pdef} we obtain
\begin{align}
\mathcal{F}(x; \mu, \Sigma_{q_{\theta}})
=&   \; \delta\Big( \frac{X_1+X_2-\mu_1-\mu_2 }{\sqrt{2}} \Big)\nonumber  \\
& \times  \tilde{\mathcal{N}}\Big( (-X_1+X_2+\mu_1-\mu_2 ) / \sqrt{2}; 0, \sqrt{2} \sigma \Big) \,  \nonumber \\
=& \sqrt{2} \times \delta\Big( X_1+X_2-1 \Big) \nonumber \\ & \times \tilde{\mathcal{N}}\Big(  \sqrt{2}  (-X_1+\mu_1 ) ; 0, \sqrt{2} \sigma \Big)  \nonumber \\
=&  \delta\Big( X_1+X_2-1 \Big) \times \tilde{\mathcal{N}}\Big(  X_1; \mu_1  \sigma \Big) \, .
\end{align}
Using these relations it is then possible to verify that the marginal probabilities are
\begin{align}
&p(x_1) \equiv \int dx_2 \; \mathcal{F}(x; \mu, \Sigma_{q_{\theta}})  = \tilde{\mathcal{N}}\Big(  x_1; \mu_1 , \sigma \Big) \, ,  \\
&p(x_2) \equiv \int dx_1 \; \mathcal{F}(x; \mu, \Sigma_{q_{\theta}}) = \tilde{\mathcal{N}}\Big(  x_2; \mu_2 , \sigma \Big) \, ,
\end{align}
while the cumulative distribution functions satisfy
\begin{equation}
P(x_1) \equiv \int_{0}^{x_1} \; dx_1^{\prime}\, p(x_1^{\prime}) = 1-P(x_2) \, .
\end{equation}
%
\section{Generating matter power spectra}\label{app:matpk}

In order to obtain the training, validation and test data, we use the recently developed code {\tt ReACT} \cite{Bose:2020wch} which calculates modified power spectra using the halo-model reaction method developed in Ref.~\cite{Cataneo:2018cic}. %
This method has been shown to be accurate to $\approx 2\%$ up to $k\sim 2.5 ~h/{\rm Mpc}$ at $z\leq1$ when compared to full N-body simulations in $f(R)$, DGP and $w$CDM models (see Figs. 3, 8 and 10 in Ref.~\cite{Cataneo:2018cic}). 
In this paper we model power spectra as described in Ref.~\cite{Cataneo:2018cic} with the exception that the pseudo power spectrum is modelled using the halofit formula of Ref.~\cite{Takahashi:2012em} which has the same level of accuracy as the approach used in Ref.~\cite{Cataneo:2018cic}. 
We refer the reader to Refs.~\cite{Cataneo:2014kaa,Bose:2020wch} for more details on the generation of the training data. 
Power spectra are generated in four redshift bins $z\in\{1.5,0.785,0.478,0.1\}$ and one hundred $k$-bins in the range $0.01 \leq k \leq 2.5 ~ h/{\rm Mpc}$ at equal intervals in log-space.
These binning choices are made according to that expected from a Euclid-like survey \cite{Laureijs:2011gra,Blanchard:2019oqi}. 
The maximum cutoff in $k$ is chosen to maintain a $\sim 2\%$ accuracy between the power spectrum generated from $\tt{ReACT}$ and simulations as shown in Ref.~\cite{Cataneo:2018cic}.
Each power spectrum is then generated by sampling the parameter space defining each model and passing these values to {\tt ReACT}. 
The $\Lambda$CDM parameter space is sampled using a Gaussian distribution centered on the Planck 2018 best fit parameters \cite{Aghanim:2018eyx}, with each standard deviation given by the Euclid `pessimistic' forecast results using weak lensing plus spectroscopic galaxy clustering (WL+${\rm GC}_{\rm s}$) \cite{Blanchard:2019oqi}. 
For $f(R)$, DGP and $w$CDM we use a Gaussian centered at the values which are equivalent to $\Lambda$CDM, namely~\footnote{For model parameter definitions we refer the reader to App.~C of Ref.~\cite{Bose:2020wch}.} $f_{\rm R0} = \Omega_{\rm rc} = w_a = 0$ and $w_0=-1$.
The standard deviations for $f(R)$ and DGP parameters are given by the recent results of Ref.~\cite{Bose:2020wch} and are summarised in Table~\ref{tab:params}.
The standard deviation for the $w$CDM parameters are taken again from the pessimistic WL+${\rm GC}_{\rm s}$ forecasts for Euclid of Ref.~\cite{Blanchard:2019oqi}. 
%

%
\section{Generating `random' power spectra}\label{app:randpk}
In this Appendix we describe how we generate power spectra with random features representing potentially exotic extensions to $\Lambda$CDM not encompassed by any of the $w$CDM, $f(R)$ or DGP models. 
The random features will be encoded in a filter in the form of a $100 \times 4$ array such that each filter has the same dimension as each example in the training set. 
We denote the $i^{\rm th}$ row and $j^{\rm th}$ column filter entry as $F[i,j]$, where $i$ corresponds to a $k \in [0.01,2.5] h/{\rm Mpc}$ and $j$ to a $z\in\{1.5,0.785,0.478,0.1 \}$. 
$F[i,j]$ assumes values centered around $1$ with a value of $1$ indicating no modification in that $k$ and $z$ bin. 
This filter can then be applied to an example from our training set, $P_{\rm ref}[i,j]$ to obtain a randomly modified power spectrum
\begin{equation}
    P_{\rm random}[i,j] = F[i,j] \times P_{\rm ref}[i,j]. \label{eq:filtertopk}
\end{equation}
The filter $F[i,j]$ is constructed in the following manner: 
\begin{enumerate}
    \item 
    Randomly select an $i_{\rm 0} \in [1,100]$ and $j_{\rm 0} \in[1,4]$.
    \item
    Assign $F[i_{0},j_{0}] = 1 + \Delta k \times  R(-1,1)$ where $R(-1,1)$ denotes a random real value between $-1$ and $1$. 
    \item
    For all integers $j \in [1,4] $, assign $F[i_{0},j\pm 1] = F[i_0,j] + \Delta z \times R(-1,1)$, starting with $j_{0}$. 
    \item
    For all integers $j \in [1,4]$, assign $F[i_0 \pm 1, j] = F[i_0 , j] + \delta k \times R(-1,1)$.  
    \item
    Repeat steps (3) and (4) for all $i\in[1,100]$ starting with $i_0$. 
\end{enumerate}
The quantities $\Delta k$, $\Delta z$ and $\delta k$ denote the maximum initial modification, the maximum difference between neighbouring columns and the maximum difference between neighbouring rows. 
These are free parameters which we set to $\Delta k = 0.1$, $\Delta z = 0.2$, and $\delta k = 0.005$. 
Our initial modification can therefore be no more than $10\%$, neighbouring $z$ points cannot vary by more than $20\%$ and neighbouring $k$ points cannot vary by more than $0.5\%$. 

Steps (1)-(5) alone generate a filter that is very noisy since there is no smoothing. %
We thus apply a further step that averages each entry along the $k-$direction over a bin of width $N_k$ \footnote{Note that we pad the edges of the array with additional values using steps (3) and (4) to smooth the boundaries.}.
\begin{enumerate}
\setcounter{enumi}{5}
\item
    For all $j \in [1,4]$ and $i\in[1,100]$, assign $F[i,j] = \frac{1}{N_k} \sum_{m=0}^{m=N_k} F[i-\frac{N_k}{2} + m,j]$.
\end{enumerate}
Step (6) is then repeated $N_s$ times to further smooth the filter. 
This leaves us with two additional free parameters, the bin width $N_k$ and the smoothing step iterations $N_s$. 
Changing each of these parameters alters the scale in the $k$-direction of the induced features.
\begin{figure*}
\centering
\resizebox{0.5\textwidth}{!}{
\includegraphics{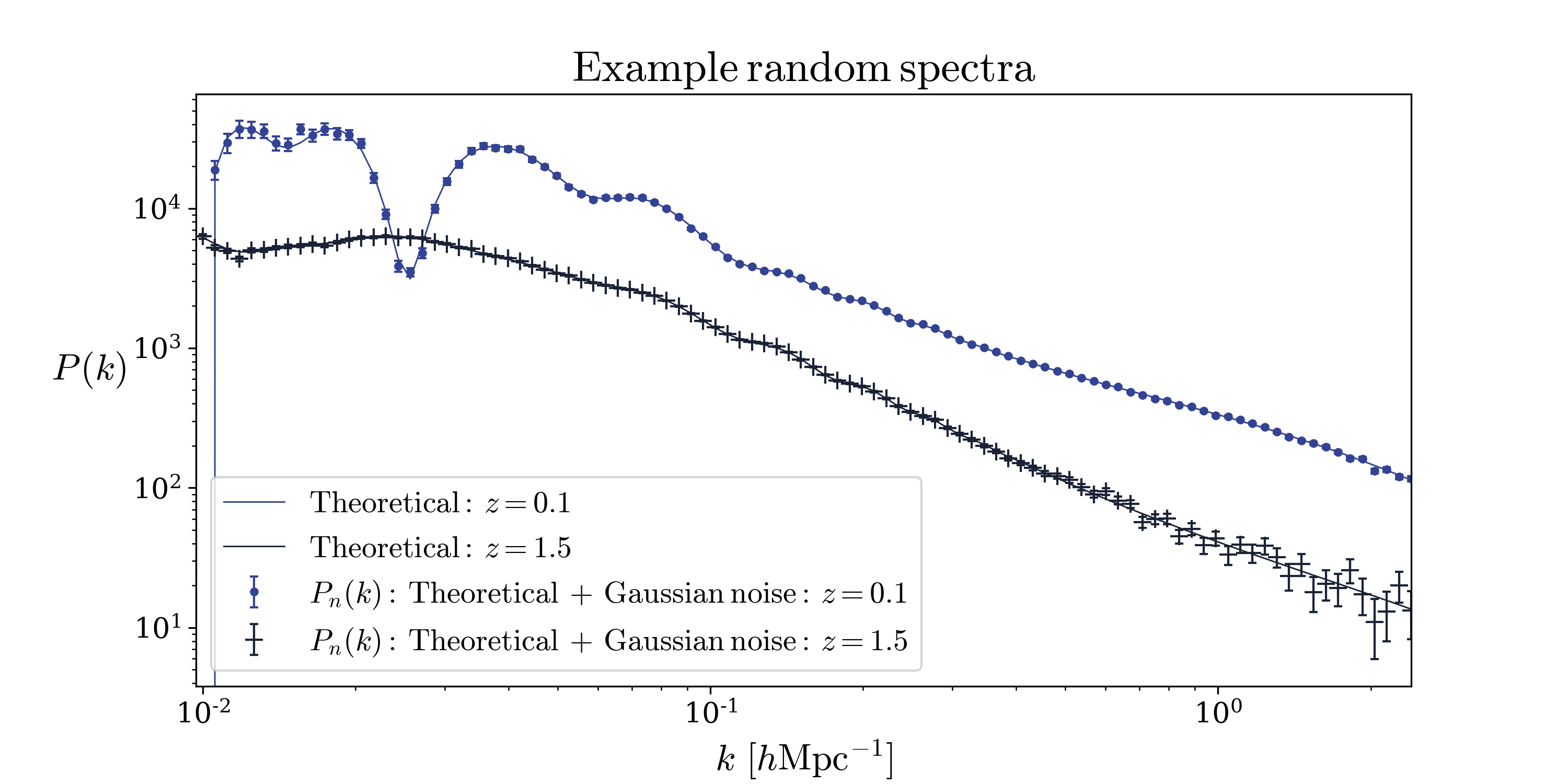}
}
\caption{Example of randomly generated matter power spectra generated using the algorithm outlined in App.~\ref{app:randpk}.
One can see that, while deviations at high $k$ remain small, large deviations are produced at low $k$ due to the poorer constraining power at these scales. 
The presence of such large deviations generally allows the five-label and two-label BNN to classify the majority of such spectra at high confidence, enabling it to pick up potential exotic theories than do not fall within the other considered classes in the training set.
}
\label{fig:random_spectra}
\end{figure*}
Finally, the maximum deviation from one in each component of the filter is then given by
\begin{align}
    \delta_M = \Delta k &+ {\rm Max}[|4-j_{0}|,|j_0-4|]\times \Delta z \nonumber \\ & +   {\rm Max}[|100-i_{0}|,|i_0-100|]\times \delta k.
\end{align}
With the selected values for $\Delta k$, $\Delta z$ and $\delta k$, this means that we can have $\delta_M = 25.65$ which corresponds to modifying a power spectrum by $2500\%$. 
Any such modification would naturally already be ruled out by observations at extremely high confidence. 
In order to moderate the imposed modifications in order that they are more likely to be consistent with current constraints we follow an additional procedure.
We begin by defining a new filter $\hat{F}[i,j]$ which conforms at some level with current observations of the power spectrum
\begin{equation}
 \hat{F}[i,j] = \epsilon[i,j] \left( F[i,j]- 1\right) + 1 \,, \label{eq:newfilt}
\end{equation}
where $\epsilon[i,j]$ is an array we shall calculate given some observational constraints. 
In order for a modification around one to be consistent with an observation at a certain confidence level we use the reduced $\chi^2$ statistic, $\chi_{\rm red}^2$, defined as 
\begin{equation}
    \chi_{\rm red}^2 = \frac{1}{N_\nu} \sum_{i=1}^{N_\nu} \frac{[P_{\rm random}(k_i) -P_{\rm ref}(k_i)]^2}{\sigma^2(k_i)}, \label{eq:chi2}
\end{equation}
where $P_{\rm random}$ and $P_{\rm ref}$ are the randomly modified and unmodified power spectra at scale $k_i$. 
The quantity $N_\nu$ is the number of degrees of freedom which in this case is the number of $k$ data points so that $N_\nu=100$ 
Finally, $\sigma(k_i)$ is the error on the $i^{\rm th}$ $k$ data point. 
To represent \emph{current} knowledge about the LSS, we model these errors based on recently completed surveys such as the BOSS survey \footnote{\protect\url{http://www.sdss3.org/}}. 
Assuming the errors are Gaussian we have \cite{Seo:2007ns}
\begin{equation}
        \sigma_p(k) =  \sqrt{\frac{4 \pi^2}{k^2 \Delta k V(z)} \times \left( P_{\rm ref}(k) + \frac{1}{\bar{n}(z)} \right)^2 +  \sigma_{sys}^2}  \, ,
        \label{eq:gaussianerrs}
\end{equation}
where $\Delta k$ is the separation between $k$ data points and $P_{\rm ref}$ is taken to be the halofit nonlinear spectrum with the Planck 2018 best fit parameters \cite{Aghanim:2018eyx}.
We take $V_{\rm eff}=1.27 ~{\rm Gpc}^3/h^3$ and $\bar{n} = 5 \times 10^{-4} ~h^3/{\rm Mpc}^3$ as an approximate volume and number density of tracers for the BOSS survey \cite{Cuesta:2015mqa}. 
We also take $\sigma_{\rm sys}^2 = 25 ~{\rm Mpc}^6/h^6$ as our modelling systematic error as in the main text (see Sec.~\ref{sec:training}).
Note we do not use the future survey specifications because these modifications should be within current constraints.
Substituting Eq.~\eqref{eq:gaussianerrs} into Eq.~\eqref{eq:chi2} and using the definition of the filter in Eq.~\eqref{eq:filtertopk} we get
\begin{equation}
\chi^2_{\rm red}(j) = \frac{1}{N_\nu} \sum_{i=1}^{N_\nu} \frac{(F[i,j]-1)^2 P_{\rm ref} [i,j]^2 \epsilon[i,j]^2}{\sigma_p^2(k_i)} \, , \label{chi2forrand} 
\end{equation}
where we perform the calculation for a fixed redshift $z_j$.
Taking the ansatz $\epsilon[i,j] = p(j) \sigma(k_i)$ we can write the unknown $p(j)$ as
\begin{equation}
    p(j) = \left[ \frac{ \chi^2_{\rm red} N_\nu}{\sum_{i=1}^{N_\nu} (F[i,j]-1)^2  P_{\rm ref}[i,j]^2 } \right]^{\frac{1}{2}} \, .
\end{equation}
Once we specify the level of deviation we want from the true spectrum in terms of the $\chi^2_{\rm red}$, we can calculate $p(j)$ for all $j\in[1,4]$. 
Using Eq.~\eqref{eq:gaussianerrs} and Eq.~\eqref{eq:newfilt}, we can then obtain our constrained random spectrum array (see Eq.~\ref{eq:filtertopk}). 
We select $\chi^2_{\rm red}\in[1,3]$ to represent modifications which deviate on average up to $\sim 3\sigma$ from the reference spectrum.

Finally, to construct the filters, we assume $P_{\rm ref} = P_{\rm pl}$, i.e.~a halofit generated power spectra with the Planck 2018 best fit cosmology (see Table~\ref{tab:params}). 
To enhance the randomness of the random spectra, we then apply the filter to a randomly selected reference spectrum in the training data set, i.e.~from the classes $f(R)$, DGP, $w$CDM and $\Lambda$CDM. 
This assumes the approximation $P_{\rm ref} \approx P_{\rm pl}$ in Eq.~\ref{chi2forrand}. 
We have a freedom to choose the level of deviation of the random spectra from Planck and we produce many thousand random spectra, making this approximation not of significant consequence to our final results.
Further, recall our reference spectra are chosen within a Gaussian distribution centered about the Planck best fits with standard deviations using recent Euclid forecasts \cite{Aghanim:2018eyx,Blanchard:2019oqi}.
Examples of the shape of random power spectra generated using this algorithm can be seen in Fig.~\ref{fig:random_spectra}. 

%
\section{Comparison with MCMC}\label{app:MCMC}

In this Appendix we elaborate further on the comparison and interplay among a classifier based on a BNN such as the one presented in this work and  the widely used statistical inference method of MCMC. 
%
We already discussed in the main text the fact that a fair comparison between the two methods presents some significant issues, and the related reasons.
One may still wonder if an MCMC analysis of mock data such as those used for our classification examples would be able to correctly identify the underlying theory and yield the corresponding bayesian evidence, should one be willing to pay the corresponding much higher computational cost. It turns out that this is not so straightforward.
%
%
%
The first issue is that the Gaussian errors associated with the data vector $P(k;z)$ are minute at small scales, making this analysis very sensitive to even the smallest systematic modelling errors. 
We introduced $\sigma_{\rm sys} $ in Eq.~\eqref{eq:Euclid_error} for this very reason. 
Nevertheless, this error was chosen for a very particular model and data vector, namely $\Lambda$CDM with a Planck cosmology, and it is unclear how much modelling inaccuracies and noise would bias other MCMC analyses.
Indeed, we explicitly checked this issue by running MCMCs on the examples discussed in Sec.~\ref{sec:example}. 
The resulting constraints show significant biases that make any attempt to compute a Bayesian evidence meaningless. 
A meaningful result could only be obtained by increasing  $\sigma_{\rm sys}$ until the final contours include the fiducial values, a process that is clearly unjustified and not applicable when considering observational data where the true value is not known. 
These issues indicate the need of a thorough investigation that goes beyond the scope of this paper.
Furthermore, in a realistic context where one does not know the actual underlying theory, in order to run an MCMC analysis to constrain extensions to $\Lambda$CDM it is necessary to choose an appropriate parameterisation which picks up the modification via a deviation from  the fiducial parameter values.
Typically generic parameters are proposed such as the growth-index which serve to account for the presence of physics beyond $\Lambda$CDM should they deviate from their fiducial value. 
A deviation in a non-$\Lambda$CDM parameter could result in significant biases in the standard $\Lambda$CDM cosmological parameters.
Classifiers such as those considered in this work possess an advantage over MCMC in this regard as they do not rely on a well chosen parameterisation to detect deviations, at least at test time.
This feature is particularly interesting if we consider the possibility that signatures of new physics may be present in the power spectrum that do not come from any theory for which numerical codes are available. 
Detecting such deviations with MCMCs may not be possible in the absence of accurate modelling, while the method described in this paper would still be able to provide hints of deviations from $\Lambda$CDM. 


\end{document}